\documentclass[journal=jctcce,manuscript=article, layout=traditional]{achemso}

\usepackage[utf8]{inputenc}
\usepackage{amsmath, amssymb}
\usepackage{graphicx}
\usepackage{color}
\usepackage{comment}
\usepackage{url}
\usepackage[breaklinks=true]{hyperref}
\usepackage[italicdiff]{physics}
\usepackage[version=4]{mhchem}
\usepackage{bm}
\usepackage{multirow}
\usepackage{mciteplus}
\usepackage{chemscheme}

\setkeys{acs}{abbreviations=true}

\newcommand{\mr}[1]{\mathrm{#1}}
\newcommand{\mc}[1]{\mathcal{#1}}
\newcommand{\wt}[1]{\widetilde{#1}}

\newcommand{\bmkp}{\bm{\kappa}}
\newcommand{\bmth}{\bm{\theta}}

\newcommand{\bmf}{\bm{F}}
\newcommand{\pol}{polarizability }

\date{\today}

\title{Analytical formulation of the second-order derivative of energy for orbital-optimized variational quantum eigensolver: application to polarizability}

\author{Yuya O. Nakagawa}
\email{nakagawa@qunasys.com}
\author{Jiabao Chen}
\affiliation{QunaSys Inc., Aqua Hakusan Building 9F, 1-13-7 Hakusan, Bunkyo, Tokyo 113-0001, Japan}

\author{Shotaro Sudo}
\author{Yu-ya Ohnishi}
\affiliation{Materials Informatics Initiative, RD Technology and Digital Transformation Center,
JSR Corporation, 3-103-9, Tonomachi, Kawasaki-ku,
Kawasaki, Kanagawa, 210-0821, Japan}

\author{Wataru Mizukami}
\affiliation{Graduate School of Engineering Science, Osaka University, 1-3 Machikaneyama, Toyonaka, Osaka 560-8531, Japan}
\alsoaffiliation{Center for Quantum Information and Quantum Biology, Osaka University, Japan}
\alsoaffiliation{JST, PRESTO, 4-1-8 Honcho, Kawaguchi, Saitama 332-0012, Japan}

\begin{document}

\begin{abstract}
We develop a quantum-classical hybrid algorithm to calculate the analytical second-order derivative of the energy for the orbital-optimized variational quantum eigensolver (OO-VQE), which is a method to calculate eigenenergies of a given molecular Hamiltonian by utilizing near-term quantum computers and classical computers.
We show that all quantities required in the algorithm to calculate the derivative can be evaluated on quantum computers as standard quantum expectation values without using any ancillary qubits.
We validate our formula by numerical simulations of quantum circuits for computing the \pol of the water molecule, which is the second-order derivative of the energy with respect to the electric field.
Moreover, the polarizabilities and refractive indices of thiophene and furan molecules are calculated as a testbed for possible industrial applications. 
We finally analyze the error-scaling of the estimated polarizabilities obtained by the proposed analytical derivative versus the numerical one obtained by the finite difference.
Numerical calculations suggest that our analytical derivative requires fewer measurements (runs) on quantum computers than the numerical derivative to achieve the same fixed accuracy.
\end{abstract}

\maketitle

%%%%% ----- %%%%%
\section{Introduction}
With significant developments in quantum technology, quantum computers~\cite{nielsen2002quantum} have now been realized in their primitive forms.
Such quantum computers are called noisy intermediate-scale quantum (NISQ) devices~\cite{Preskill2018}, which are composed of dozens to thousands of quantum bits (qubits) without quantum error correction.
NISQ devices have now been shown to surpass classical computers in a specific computational task with no practical application~\cite{Arute2019, Wu2021, Zhong1460, Madsen2022}.
The noisy nature of NISQ devices prevents us from executing deep quantum circuits, which are often required in quantum algorithms that are expected to have exponential speedup over classical algorithms, such as Shor's prime factorization, Grover's search, and quantum phase estimation~\cite{shor1999polynomial, grover1996fast, kitaev1995quantum, Cleve1998}.
Still, a lot of effort has been put into utilizing such devices for practical purposes.

Quantum chemistry calculation has been considered one of the most promising fields for the application of NISQ devices.
A central algorithm for such applications is the variational quantum eigensolver (VQE)~\cite{peruzzo2014variational, mcclean2016theory, tilly2021variational}, which can calculate approximate eigenvalues of a given molecular Hamiltonian by using the variational principle of quantum mechanics.
In VQE, a quantum computer realizes a quantum state corresponding to a wavefunction for a given Hamiltonian and outputs the expectation value of the Hamiltonian (i.e., energy), while a classical computer gives an instruction to the quantum computer for minimizing the expectation value.  
This quantum-classical hybrid architecture allows us to use only shallow quantum circuits that may be executable on NISQ devices, and there were already plenty of experiments performing VQE for small quantum systems (molecules)~\cite{peruzzo2014variational,kandala2017hardware, colless2018computation, kandala2019error}.
Since quantum states on quantum computers can express superposition of the exponentially large number of configurations with the number of qubits, VQE may outperform the classical computation of the molecular energy~\cite{mcardle2018quantum, Cao2018}.

In quantum chemistry calculations, energy is one of the most important quantities, but the derivative of the energy with respect to external parameters (here we call it the ``energy derivative") is also indispensable to predicting the properties of materials.
Although the primary purpose of VQE is to find eigenenergies of a given Hamiltonian, there are several studies extending the VQE-based algorithms to calculate the energy derivative~\cite{mitarai2020theory, parrish2019hybrid, tamiya2021, parrish2021analytical, yalouz2022analytical, omiya2022,hohenstein2022efficient} (see also Refs.~\citenum{o2019calculating, o2021efficient} for proposals not based on VQE). 
Most of those studies treated the energy derivative with respect to the atomic coordinates of molecules under the Born-Oppenheimer approximation, which gives force acting on the molecule or the vibrational frequency of the molecule. 

However, there is another important derivative: the derivative with respect to electromagnetic fields (i.e., the ``electromagnetic derivative").
The electromagnetic derivative involves various crucial properties of molecules, especially optical properties; for example, the \pol is related to the refractive index of materials, the infrared (IR) and Raman intensities are clues to specify molecules, and the circular dichroism is used to detect the chirality of molecules.
Therefore, it is demanded to develop a method to calculate the electromagnetic derivative of energy with quantum computers.

In this study, we develop a quantum-classical hybrid algorithm to calculate the second-order derivative of energy obtained by the method called orbital-optimized VQE (OO-VQE)~\cite{Takeshita2020, Mizukami2020, Sokolov2020JCP}.
OO-VQE is an extension of VQE that can save quantum computational resources compared with the usual VQE by leveraging orbital optimization on classical computers.
We derive a formula for the analytical derivative of energy based on the variational condition of OO-VQE, which can avoid the problem of discretization error in the numerical derivative using the finite difference method.
We apply the formula of the analytical derivative to the polarizability, which is the second-order derivative of energy with respect to the electric field.
We also discuss the potential advantages of our analytical derivative over the numerical derivative by analyzing the error-scaling of the estimated \pol and performing numerical simulations of quantum circuits with fluctuating outputs.
Our work expands the range of applications of quantum computers, especially NISQ devices, in quantum chemistry.

This paper is organized as follows.
In Sec.~\ref{sec: preliminaries}, we explain several preliminary techniques to derive our formula of the analytical derivative.
Section~\ref{sec: general theory} provides an algorithm to calculate the analytical second-order derivative of OO-VQE energy for general external parameters.
We apply the algorithm to the calculation of the \pol and validate it by numerical simulation in Sec.~\ref{sec: pol application}.
Finally, we analyze the error-scaling of the estimated \pol and compare the computational cost for quantum computers between the analytical derivative and the numerical derivative in Sec.~\ref{sec: error analysis}
Conclusion and outlook are discussed in Sec.~\ref{sec: summary}.

%%%%% ----- %%%%%
\section{Preliminaries \label{sec: preliminaries}}
In this section, we review an algorithm of OO-VQE and several techniques to derive our main results in Secs.~\ref{sec: general theory} and \ref{sec: pol application}.   

\subsection{Notations}
We first define the notations of several quantities.
We consider the second quantized Hamiltonian for electronic states of a given molecule,
\begin{equation}
\label{eq: full-space hamiltonian}
\begin{split}
 \hat{H}_{\mr{full}}(\bmf, \bmkp) &\equiv E_\mr{c}(\bmf) + \sum_{p,q=1}^{N_\mathrm{MO}} \sum_{\sigma=\uparrow, \downarrow} h_{pq}(\bmf, \bmkp)\hat{a}_{p\sigma}^\dag\hat{a}_{q\sigma}
 \\ &+ \frac{1}{2} \sum_{p,q,r,s=1}^{N_\mathrm{MO}} \sum_{\sigma,\tau=\uparrow, \downarrow} g_{pqrs}(\bmf, \bmkp)\hat{a}^\dag_{p\sigma}\hat{a}^\dag_{q\tau}\hat{a}_{r\tau}\hat{a}_{s\sigma},
\end{split}
\end{equation}
where $\bmf$ are external parameters such as electric fields applied on the molecule, $\bmkp$ are the orbital parameters that determine the molecular orbitals (MOs), and $N_\mathrm{MO}$ is the number of MOs.
$E_\mr{c}(\bmf)$, $h_{pq}(\bmf, \bmkp)$, and $g_{pqrs}(\bmf, \bmkp)$ are scalars calculated by the classical computers (e.g., molecular integrals\footnote{
Note that we follow Ref.~\citenum{mcardle2018quantum} for the order of the indices of $g_{pqrs}$.
Explicit expressions for $E_\mr{c}(\bmf), h_{pq}(\bmf, \bmkp)$, and $g_{pqrs}(\bmf, \bmkp)$ for the case of the \pol are given in Sec.~\ref{sec: pol application}.}),
and  $\hat{a}_{p\sigma}\left(\hat{a}^\dag_{p\sigma}\right)$ is an annihilation (creation) operator corresponding to $p$th MO with spin $\sigma=\uparrow, \downarrow$.
The annihilation and creation operators satisfy the fermionic anti-commutation relation, $\left\{\hat{a}_{p\sigma},\hat{a}_{q\sigma'}\right\} = \left\{\hat{a}^\dag_{p\sigma},\hat{a}^\dag_{q\sigma'}\right\} = 0$ and $\left\{\hat{a}_{p\sigma},\hat{a}^\dag_{q\sigma'}\right\}=\delta_{pq}\delta_{\sigma\sigma'}$,
where $\left\{A, B\right\}\equiv AB+BA$ and $\delta$ is the Kronecker delta. 

We apply the active space approximation to this Hamiltonian in OO-VQE.
The orbitals are divided into three groups: the core (doubly-occupied) orbitals, the active orbitals, and the virtual orbitals.
The wavefunction under the active space approximation is written as $\ket{\Psi} = \ket{\mr{vac}}_\mr{vir} \otimes \ket{\psi} \otimes\ket{\uparrow\downarrow}_\mr{core}$, where $\ket{\mr{vac}}_\mr{vir}$ is the vacuum (vacant) state for the virtual orbitals, $\ket{\psi}$ is a wavefunction in the active space, and $\ket{\uparrow\downarrow}_\mr{core}$ is the doubly-occupied state for the core orbitals.
The full-orbital Hamiltonian $\hat{H}_\mr{full}$ is projected onto the effective Hamiltonian $\hat{H}$ in the active space by the projector $\mc{P}$ as 
\begin{align}
 \mc{P} \equiv \left( \ket{\mr{vac}}_\mr{vir} \bra{\mr{vac}}_\mr{vir} \right) \otimes \hat{I}_{AS} \otimes \left( \ket{\uparrow\downarrow}_\mr{core} \bra{\uparrow\downarrow}_\mr{core} \right), \\
 \mc{P} \hat{H}_\mr{full}(\bmf, \bmkp) \mc{P} = \hat{H}(\bmf, \bmkp),
\end{align}
where $\hat{I}_{AS}$ is the identify operator in the active space.
The projected Hamiltonian $\hat{H}(\bmf, \bmkp)$ has still the same form of $\hat{H}_\mr{full}$,
\begin{align}
\begin{split}
 \hat{H}(\bmf, \bmkp) &\equiv E'_\mr{c}(\bmf) + \sum_{p,q}' \sum_{\sigma=\uparrow, \downarrow} h'_{pq}(\bmf, \bmkp) \hat{a}_{p\sigma}^\dag\hat{a}_{q\sigma}
 \\ &+ \frac{1}{2} \sum_{pqrs}' \sum_{\sigma,\tau=\uparrow, \downarrow} g'_{pqrs}(\bmf, \bmkp)\hat{a}^\dag_{p\sigma}\hat{a}^\dag_{q\tau}\hat{a}_{r\tau}\hat{a}_{s\sigma},
\end{split}
\end{align}
where the summation runs only for the active space orbitals.
The values of $E'(\bmf), h'_{pq}(\bmf, \bmkp)$ and $g'_{pqrs}(\bmf, \bmkp)$ are again calculated by classical computers efficiently (i.e., polynomial classical computation cost in the number of MOs).

\subsection{Review of orbital-optimized variational quantum eigensolver (OO-VQE)}
Next, we review the algorithm of OO-VQE~\cite{Takeshita2020,Mizukami2020,Sokolov2020JCP}.
OO-VQE is an algorithm to calculate eigenstates and eigenenergies of a given molecular Hamiltonian under the active space approximation by optimizing the electron configuration and the MOs simultaneously.
The counterpart of OO-VQE in classical computational methods is the multiconfigurational self-consistent field (MCSCF)~\cite{Szalay2012, Roos2016multiconfigurational}.
In OO-VQE, we optimize two kinds of parameters to find a better description of the electronic state of the molecule: orbital parameters $\bmkp$ and circuit parameters $\bmth$.
The orbital parameters dictate the orbital rotation among the MOs and each component of them is specified by distinct pairs of the MOs. For example, $\kappa_{pq}$ determines the orbital rotation between the $p$th and $q$th MOs.
The action of the orbital parameters is expressed as
\begin{equation} \label{eq: def OO rotation}
\begin{split}
  \hat{U}_{\mr{OO}}(\bmkp) \equiv \exp\left[ \sum_{p<q} \sum_{\sigma=\uparrow,\downarrow} \kappa_{pq}\left(\hat{a}_{p\sigma}^\dag\hat{a}_{q\sigma} - \hat{a}_{q\sigma}^\dag\hat{a}_{p\sigma}\right)\right], \\
  \hat{H}_\mr{full}(\bmf, \bmkp) = \hat{U}_{\mr{OO}}(\bmkp)^\dag \hat{H}_\mr{full}(\bmf, \bmkp=\bm{0}) \hat{U}_{\mr{OO}}(\bmkp).
\end{split}
\end{equation}
It should be noted that the orbital rotation is performed solely with classical computers; the orbital rotation can be viewed as the transformation of the coefficients of molecular orbitals expanded by atomic orbitals~\cite{Bozkaya2011}, and $\hat{H}_\mr{full}(\bmf, \bmkp)$ can be calculated from $\hat{H}_\mr{full}(\bmf, \bmkp=\bm{0})$ with classical computers.
On the other hand, the circuit parameters $\bmth =(\theta_1, \cdots, \theta_M)$, where $M$ is the number of them, are defined through an {\it ansatz} (a trial quantum state) in the active space,
\begin{equation} \label{eq: def circ-params}
\ket{\psi(\bmth)} = \hat{U}(\bmth)\ket{\psi_{\rm{ref}}},
\end{equation}
where $\hat{U}(\bmth)$ is a unitary operator parameterized by $\bmth$ and $\ket{\psi_{\rm{ref}}}$ is the reference state such as the Hartree-Fock state.
When the active space consists of $N_{AS}$ spatial MOs, we can prepare the state $\ket{\psi(\bmth)}$ as a quantum state on a quantum computer of $2N_{AS}$ quantum bits, or qubits.
The unitary $\hat{U}(\bmth)$ is typically defined as a specific quantum circuit on a quantum computer and the parameters $\bmth$ are set to rotational angles for the gates included in the quantum circuit
(see, e.g., Refs.~\citenum{mcardle2018quantum, Cao2018} for a review of the applications of quantum computers to quantum chemistry).

OO-VQE is a variational method, and the cost function to be minimized in OO-VQE is the energy expectation value of the trial state $\ket{\psi(\bmth)}$, 
\begin{equation}
 E(\bmf, \bmkp, \bmth) \equiv \ev{\hat{H}(\bmf, \bmkp)}{\psi(\bmth)},
\end{equation}
where $\hat{H}(\bmf, \bmkp) = \mc{P}\hat{H}_\mr{full}(\bmf, \bmkp)\mc{P}$ is the projected Hamiltonian in the active space.
The OO-VQE algorithm updates the parameters $\bmkp$ and $\bmth$ iteratively.
First, for some fixed $\bmkp_0$, the optimization of the circuit parameters $\bmth$ is performed by evaluating $E(\bmf, \bmkp_0, \bmth)$ on quantum computers.
This part is performed in the same manner as the conventional VQE algorithm~\cite{peruzzo2014variational, mcclean2016theory, tilly2021variational}; the circuit parameters $\bmth$ are iteratively updated by classical optimization algorithms with monitoring the value of $E(\bmf, \bmkp_0, \bmth)$.
Second, the optimized circuit parameters $\bmth_0$ for the fixed $\bmkp_0$ are then used to evaluate the one-particle and two-particle reduced density matrices (1,2-RDMs), 
\begin{equation} \label{eq: def of RDMs}
\rho^{(1)}_{pq} \equiv \ev{\sum_\sigma \hat{a}_{p\sigma}^\dag\hat{a}_{q\sigma}}{\Psi},\;
\rho^{(2)}_{pqrs}(\bmth_0) \equiv \ev{\sum_{\sigma\tau} \hat{a}_{p\sigma}^\dag \hat{a}_{q\tau}^\dag \hat{a}_{r\tau} \hat{a}_{s\sigma}}{\Psi},
\end{equation}
where $\ket{\Psi}= \ket{\mr{vac}}_\mr{vir} \otimes \ket{\psi(\bmth)} \otimes\ket{\uparrow\downarrow}_\mr{core}$ is a quantum state in the whole space
\footnote{
It should be stressed that only quantum computers of $2N_{AS}$ qubits are needed to evaluate those RDMs by projecting the operators $\hat{a}_{p\sigma}^\dag \hat{a}_{q\sigma}$ and $\hat{a}_{p\sigma}^\dag \hat{a}_{q\tau}^\dag \hat{a}_{r\tau} \hat{a}_{s\sigma}$ onto the active space and measuring the expectation values of them for $\ket{\psi(\bmth)})$.}.
As we will see in the next subsection, the partial derivative of $E(\bmf, \bmkp, \bmth)$ with respect to $\bmkp$ can be computed by the values of 1,2-RDMs.
The orbital parameters $\bmkp_0$ are updated to $\bmkp_1$ to lower the value of the cost function by using that derivative.
One repeats these procedures to update $\bmkp$ and $\bmth$ until the value of the cost function (or energy) converges, and finally obtains the output as the optimized energy of OO-VQE:
\begin{equation}
 E^*(\bmf) \equiv E(\bmf, \bmkp^*(\bmf), \bmth^*(\bmf)),
\end{equation}
where we denote the optimal parameters obtained by OO-VQE $\bmkp^*(\bmf)$ and $\bmth^*(\bmf)$.
The optimized energy calculated by OO-VQE is still dependent on the external parameters $\bmf$, and we aim at calculating the derivative of the optimal energy such as $\pdv{E^*(\bmf)}{\bmf}$ in this study.

\subsection{Partial derivatives of energy with respect to external, orbital, and circuit parameters}
To calculate the derivative of the optimal energy $E^*(\bmf)$ obtained by OO-VQE, the partial derivatives of the cost function $E(\bmf, \bmkp, \bmth)$ such as 
\begin{equation}
 \pdv{E(\bmf, \bmkp, \bmth)}{F_d},
 \pdv{E(\bmf, \bmkp, \bmth)}{\kappa_{pq}}, \pdv{E(\bmf, \bmkp, \bmth)}{\kappa_{pq}}{\theta_k}, \cdots,
\end{equation}
are needed.
Here we review several techniques to evaluate those partial derivatives with quantum computers.
We note that all of the techniques used to evaluate the partial derivatives were already discussed in the literature~\cite{mitarai2020theory, o2019calculating, Mizukami2020, omiya2022, Mitarai2018, Schuld2019, izmaylov2021analytic} and that the purpose of this subsection is for completeness of this article.
We first present the way to evaluate the first-order partial derivatives of $E(\bmf, \bmkp, \bmth)$ with respect to three parameters, $\bmf, \bmkp$ and $\bmth$. The cross-parameter partial derivatives such as $\pdv{E}{\kappa_{pq}}{\theta_{k}}$ can also be evaluated by combining the techniques for the corresponding parameters, and we show one such example at the end of this section.

\subsubsection{Partial derivative with respect to $F$} 
Partial derivatives of $E(\bmf, \bmkp, \bmth)$ with respect to $\bmf$ can be evaluated as ordinary expectation values for observables on quantum computers~\cite{Mitarai2018, o2019calculating}.
For example, the partial derivative $\pdv{E(\bmf, \bmkp, \bmth)}{F_d}$ is evaluated by
\begin{equation}
 \pdv{E(\bmf, \bmkp, \bmth)}{F_d} = \ev{\pdv{\hat{H}(\bmf, \bmkp)}{F_d}}{\psi(\bmth)},
\end{equation}
where $F_d$ indicates the $d$th component of the external field.
This equation means that the partial derivative can be evaluated as an expectation value for the partial derivative of the active space Hamiltonian,
$\pdv{\hat{H}(\bmf, \bmkp)}{F_d} = \mc{P}\pdv{\hat{H}_\mr{full}(\bmf, \bmkp)}{F_d}\mc{P}$.
This observable is explicitly written as (see Eq.~\eqref{eq: full-space hamiltonian}),
\begin{equation}
\label{eq: f-deriv hamiltonian}
\begin{split}
 \pdv{\hat{H}_\mr{full}(\bmf, \bmkp)}{F_d} & = \pdv{E_c(\bmf)}{F_d} + \sum_{pq, \sigma} \pdv{h_{pq}(\bmf, \bmkp)}{F_d} \hat{a}_{p\sigma}^\dag\hat{a}_{q\sigma}
 \\ &+ \frac{1}{2}\sum_{pqrs,\sigma\tau} \pdv{g_{pqrs}(\bmf, \bmkp)}{F_d} \hat{a}^\dag_{p\sigma}\hat{a}^\dag_{q\tau}\hat{a}_{r\tau}\hat{a}_{s\sigma}.
\end{split}
\end{equation}
Therefore, it is enough to know the partial derivatives of the coefficients 
\begin{equation}
\label{eq: f-deriv coeff}
\pdv{E_c(\bmf)}{F_d}, \pdv{h_{pq}(\bmf, \bmkp)}{F_d}, \pdv{g_{pqrs}(\bmf, \bmkp)}{F_d}.
\end{equation}
In most cases including the \pol which we focus on in the later sections, the values in Eq.~\eqref{eq: f-deriv coeff} can be analytically computed by classical computers in an efficient way.
We note that those partial derivatives of the coefficients do not contain contributions from the orbital response.

\subsubsection{Partial derivative with respect to $\kappa$}
Partial derivatives of $E(\bmf, \bmkp, \bmth)$ with respect to the orbital parameters $\bm{\kappa}$ can be evaluated by properly combining the values of 1,2-RDMs~\cite{Mizukami2020, omiya2022}, $\rho^{(1)}_{pq}$ and $\rho^{(2)}_{pqrs}$ [Eq.~\eqref{eq: def of RDMs}].
The following equations hold from the definition of $\hat{U}_\mr{OO}(\bm{\kappa})$ [Eq.~\eqref{eq: def OO rotation}]:
\begin{align} 
\pdv{E(\bmf, \bmkp, \bmth)}{\kappa_{pq}} \big|_{\bmkp=\bm{0}} 
&= \ev{\left[ \hat{H}_\mr{full}(\bmf, \bmkp), \hat{\kappa}_{pq} \right]}{\Psi} \big|_{\bm{\kappa}=\bm{0}},
\label{eq: kappa 1st deriv} \\
\pdv{E(\bmf, \bmkp, \bmth)}{\kappa_{pq}}{\kappa_{rs}} \big|_{\bm{\kappa}=\bm{0}} 
& = \frac{1}{2} 
\ev{\left[ \left[ \hat{H}_\mr{full}(\bmf, \bmkp), \hat{\kappa}_{rs}, \right], \hat{\kappa}_{pq} \right]}{\Psi} \big|_{\bm{\kappa}=\bm{0}} \nonumber \\
& + \frac{1}{2} 
\ev{ \left[ \left[ \hat{H}_\mr{full}(\bmf, \bmkp), \hat{\kappa}_{pq} \right], \hat{\kappa}_{rs} \right]}{\Psi} \big|_{\bm{\kappa}=\bm{0}},
\label{eq: kappa 2nd deriv}
\end{align}
where $[A,B]=AB-BA$ is a commutator, $\hat{\kappa}_{pq}$ is  
\begin{equation}
 \hat{\kappa}_{pq} = \sum_\sigma (\hat{a}^\dag_{p\sigma}\hat{a}_{q\sigma}-\hat{a}_{q\sigma}^\dag\hat{a}_{p\sigma}),
\end{equation}
and $\ket{\Psi}= \ket{\mr{vac}}_\mr{vir} \otimes \ket{\psi(\bmth)} \otimes\ket{\uparrow\downarrow}_\mr{core}$ is a quantum state in the whole space.
We can show the following equation by an explicit calculation,
\begin{align}
\begin{split}
 & \ev{\left[ \hat{H}_\mr{full}(\bmf, \bmkp), \hat{\kappa}_{pq} \right]}{\Psi} \big|_{\bm{\kappa}=\bm{0}} \\
&= \sum_r {h}_{pr}\rho^{(1)}_{rq} -\sum_r {h}_{qr}\rho_{rp}^{(1)} -\sum_r {h}_{rq}\rho_{pr}^{(1)} +\sum_r {h}_{rp}\rho_{qr}^{(1)} \\
& - \sum_{rst} g_{qrst}\rho_{prst}^{(2)} + \sum_{rst} g_{prst}\rho_{qrst}^{(2)} + \sum_{rst} g_{rspt}\rho_{rsqt}^{(2)} - \sum_{rst} g_{rsqt}\rho_{rspt}^{(2)},
\end{split}
\label{eq: [H,kappa]}
\end{align}
assuming the symmetry $h_{pq}=h_{qp}, g_{pqrs}=g_{sqrp}=g_{prqs}=g_{qpsr}$.
Equation~\eqref{eq: [H,kappa]} means that evaluating the 1,2-RDMs is enough to calculate the partial derivative of $E(\bmf,\bmkp,\bmth)$ with respect to the orbital parameters $\bmkp$.
The second-order derivative~[Eq.~\eqref{eq: kappa 2nd deriv}] can also be evaluated by combining the values of the 1,2-RDMs (see Supporting Information).

\subsubsection{Partial derivative with respect to $\theta$ \label{subsubsec: pdv_E_theta}}
Partial derivatives of $E(\bmf, \bmkp, \bmth)$ with respect to the circuit parameters $\bmth$ can also be evaluated as expectation values of proper observables on $2N_{AS}$ qubits with the technique called ``parameter shift rule"~\cite{Mitarai2018, Schuld2019, izmaylov2021analytic}.
For simplicity, we assume that the quantum circuit $U(\bmth)$ for the ansatz $\ket{\psi(\bmth)}$ has the form
\begin{equation}
\label{eq: form of ansatz}
 \begin{split}
 \hat{U}(\bmth) &=
 \prod_{k=1}^M \exp\left[ - i\frac{\theta_k}{2} \hat{P}_k \right] \equiv \hat{U}_M(\theta_M) \cdots \hat{U}_2(\theta_2) \hat{U}_1(\theta_1),
 \end{split}
\end{equation}
where 
${\hat{P}_k\in \{I, X, Y, Z\}^{\otimes N_{AS}}}$
is the multiqubit Pauli operator on $N_{AS}$ qubits satisfying $\hat{P}_k^2 = \hat{I}$.
The parameter shift rule enables us to evaluate the partial derivative of $E(\bmf,\bmkp,\bmth)$ with respect to the circuit parameter $\theta_k$ as a sum of expectation values of $E$ at ``shifted" parameters,
\begin{equation} \label{eq: theta derivative}
 \begin{split}
 \pdv{E(\bmf, \bmkp, \bmth)}{\theta_k} =
 \frac{1}{2} \ev{\hat{U}_{k,+}(\bmth)^\dag \hat{H}(\bmf, \bmkp) \hat{U}_{k,+}(\bmth)}{\psi_\mr{ref}}
 -\frac{1}{2} \ev{\hat{U}_{k,-}(\bmth)^\dag \hat{H}(\bmf, \bmkp) \hat{U}_{k,-}(\bmth)}{\psi_\mr{ref}}
 \end{split}
\end{equation}
where $\hat{U}_{k,\pm}(\bmth)$ is defined as 
\begin{equation}
 \hat{U}_{k,\pm}(\bmth) \equiv
 \left( \prod_{k' > k} \hat{U}_{k'}(\theta_{k'})\right)
 \hat{U}_k \left( \theta_k \pm \frac{\pi}{2}\right)
 \left(\prod_{k'<k} \hat{U}_{k'}(\theta_{k'})\right).
\end{equation}
This equation means that the partial derivative with respect to $\theta_k$ is evaluated by the difference between expectation values of $\hat{H}(\bmf,\bmkp)$ for two states, $\hat{U}_{k,+}(\bmth)\ket{\psi_\mr{ref}}$ and $\hat{U}_{k,-}(\bmth)\ket{\psi_\mr{ref}}$.
It is straightforward to derive similar formulas for higher-order partial derivatives. 

The parameter shift rule is advantageous in evaluating the partial derivative with respect to $\bmth$ for several reasons.
First, it requires only expectation values that can be measured without any ancillary qubit.
Second, it is expected to be more robust to the noise of current quantum computers than naive numerical differentiation.
In numerical differentiation, the partial derivative is evaluated by the difference in energy between two parameters $\theta_k$ and $\theta_k + \epsilon$ for small $\epsilon \sim 0$.
In contrast, the parameter shift rule ensures that the partial derivative can be evaluated by the difference in energy between $\theta_k + \pi/2$ and $\theta_k - \pi/2$, which is distant by $\pi$.
The energy difference is generally larger for such distant parameters, so the parameter shift rule is considered to be more stable in the presence of noise in quantum computer outputs.

\subsubsection{Example of cross-parameter partial derivative}
By combining the techniques above, it is possible to evaluate the cross-parameter partial derivatives of $E(\bmf,\bmkp,\bmth)$.
Here we show one explicit example, $\pdv{E(\bmf,\bmkp,\bmth)}{\kappa_{pq}}{\theta_k}$.
This derivative can be calculated by taking the partial derivative of Eq.~\eqref{eq: [H,kappa]} with respect to $\theta_k$, which leads to
\begin{align}
\pdv{E(\bmf, \bmkp, \bmth)}{\kappa_{pq}}{\theta_k} \big|_{\bmkp=\bm{0}} 
& = \sum_r {h}_{rq} \pdv{\rho_{pr}^{(1)}}{\theta_k} + \sum_r {h}_{rp} \pdv{\rho_{qr}^{(1)}}{\theta_k} \nonumber \\
& - \sum_{rst} g_{qrst} \pdv{\rho_{prst}^{(2)}}{\theta_k} + \sum_{rst} g_{prst} \pdv{\rho_{qrst}^{(2)}}{\theta_k} + \sum_{rst} g_{rspt} \pdv{\rho_{rsqt}^{(2)}}{\theta_k} - \sum_{rst} g_{rsqt} \pdv{\rho_{rspt}^{(2)}}{\theta_k}.
\label{eq: pdv_E_kappa_theta}
\end{align}
The partial derivatives of 1,2-RDMs with respect to $\theta_k$ are evaluated by the technique in Sec.~\ref{subsubsec: pdv_E_theta}.
By recalling that the 1,2-RDMs $\rho^{(1)}_{rs}, \rho^{(2)}_{rstu}$ are respectively expectation values of the observables $\sum_\sigma \hat{a}_{r\sigma}^\dag\hat{a}_{s\sigma}$ and $\sum_{\sigma,\tau} \hat{a}_{r\sigma}^\dag \hat{a}_{s\tau}^\dag \hat{a}_{t\tau} \hat{a}_{u\sigma}$ for the ansatz state $\ket{\psi(\bmth)}$ [see Eq.~\eqref{eq: def of RDMs}], their partial derivatives can be expressed as
\begin{align*}
 \pdv{\rho^{(1)}_{rs}}{\theta_k} & = 
 \frac{1}{2} \left( \ev{ \sum_\sigma \hat{a}_{r\sigma}^\dag\hat{a}_{s\sigma} }{\psi_{k,+}(\bmth)} 
 -\ev{ \sum_\sigma \hat{a}_{r\sigma}^\dag\hat{a}_{s\sigma} }{\psi_{k,-}(\bmth)}  \right), \\
 \pdv{\rho^{(2)}_{rstu}}{\theta_k} & = 
 \frac{1}{2} \left( \ev{ \sum_{\sigma,\tau} \hat{a}_{r\sigma}^\dag \hat{a}_{s\tau}^\dag \hat{a}_{t\tau} \hat{a}_{u\sigma} }{\psi_{k,+}(\bmth)} 
 -\ev{ \sum_{\sigma,\tau} \hat{a}_{r\sigma}^\dag \hat{a}_{s\tau}^\dag \hat{a}_{t\tau} \hat{a}_{u\sigma} }{\psi_{k,-}(\bmth)}  \right),
\end{align*}
where $\ket{\psi_{k,\pm}(\bmth)} = \hat{U}_{k,\pm}(\bmth) \ket{\psi_\mr{ref}}$.
Therefore, the partial derivatives of 1,2-RDMs with respect to $\theta_k$ are determined by expectation values for two states $\ket{\psi_{k,\pm}(\bmth)}$, and putting them into Eq.~\eqref{eq: pdv_E_kappa_theta} yields the value of the cross-parameter derivative $\pdv{E(\bmf, \bmkp, \bmth)}{\kappa_{pq}}{\theta_k}$.

%%%%% ----- %%%%%
\section{Analytical formulation of second-order derivative of OO-VQE energy \label{sec: general theory}}
In this section, we derive one of the main results of our study: an analytical formula for the second-order derivative of the energy obtained by OO-VQE.

As reviewed in the previous section, the energy $E(\bmf, \bmkp, \bmth)$ is optimal with respect to both the orbital parameters $\bmkp$ and the circuit parameters $\bmth$ after the convergence of OO-VQE: 
\begin{equation}
\label{eq: opt. cond. OO-VQE}
  \left. \pdv{E(\bmf, \bmkp, \bmth)}{\kappa_{pq}} \right|_{\bmkp=\bmkp^*(\bmf), \bmth=\bmth^*(\bmf)} = 0,
\left. \pdv{E(\bmf, \bmkp, \bmth)}{\theta_k} \right|_{\bmkp=\bmkp^*(\bmf), \bmth=\bmth^*(\bmf)} = 0,
\end{equation}
for all MO pairs $(p,q)$ and $k=1,\cdots,M$.
We are interested in the second-order derivative of the optimal energy $E^*(\bmf) = E(\bmf, \bmkp^*(\bmf), \bmth^*(\bmf))$ with respect to $\bmf$.
By using the optimal condition above, we reach the expression of the second-order derivative, 
\begin{equation}
\label{eq: formula 2nd deriv.}
 \begin{split}
 \pdv{E^*(\bmf)}{F_d}{F_{d'}} =
 \pdv{E(\bmf,\bmkp^*, \bmth^*)}{F_d}{F_{d'}}
 + \sum_{p<q} \pdv{\kappa_{pq}^*(\bmf)}{F_d} \pdv{E(\bmf, \bmkp^*, \bmth^*)}{\kappa_{pq}}{F_{d'}} \\
 + \sum_k \pdv{\theta_k^*(\bmf)}{F_d} \pdv{E(\bmf, \bmkp^*, \bmth^*)}{\theta_k}{F_{d'}}.
 \end{split}
\end{equation}
Here we use the notation such as
\begin{align}
\pdv{E(\bmf, \bmkp^*, \bmth^*)}{F_d}{ F_{d'}} \equiv
  \left. \pdv{E(\bmf, \bmkp, \bmth)}{F_d}{F_{d'}} \right|_{\bmkp=\bmkp^*(\bmf), \bmth=\bmth^*(\bmf)}, \\
  \pdv{E(\bmf, \bmkp^*,  \bmth^*)}{\kappa_{pq}}{F_{d'}} \equiv
  \left. \pdv{E(\bmf, \bmkp, \bmth)}{\kappa_{pq}}{F_{d'}} \right|_{\bmkp=\bmkp^*(\bmf), \bmth=\bmth^*(\bmf)},
\end{align}
for brevity.
We need two groups of quantities to evaluate the right hand side of Eq.~\eqref{eq: formula 2nd deriv.}.
The first group is composed of the expectation values
\begin{equation}
\label{eq: evs in 2nd deriv.}
 \pdv{E(\bmf, \bmkp^*, \bmth^*)}{F_d}{F_{d'}},
 \pdv{E(\bmf, \bmkp^*, \bmth^*)}{\kappa_{pq}}{F_{d'}},
  \pdv{E(\bmf, \bmkp^*, \bmth^*)}{\theta_k}{F_{d'}}.
\end{equation}
These quantities are partial derivatives of the OO-VQE cost function and can be evaluated by the techniques explained in Sec.~\ref{sec: preliminaries}. 
The second one is composed of the responses of the optimal parameters to the external field,
\begin{equation} \label{eq: response of optimal params}
 \pdv{\kappa_{pq}^*(\bmf)}{F_d}, \pdv{\theta_k^*(\bmf)}{F_d}.
\end{equation}
We will explain how to calculate them by combining classical and quantum computers in the following.

The responses of the optimal parameters are determined by differentiating Eq.~\eqref{eq: opt. cond. OO-VQE} by $\bmf$,
\begin{equation}
 \begin{split}
 \sum_{rs} \pdv{E(\bmf, \bmkp^*, \bmth^*)}{\kappa_{pq}}{\kappa_{rs}} \pdv{\kappa_{rs}^*(\bmf)}{F_d} + 
 \sum_l \pdv{E(\bmf, \bmkp^*, \bmth^*)}{\kappa_{pq}}{\theta_l} \pdv{\theta_l^*(\bmf)}{F_d}= - \pdv{E(\bmf, \bmkp^*, \bmth^*)}{\kappa_{pq}}{F_d}, \\
 \sum_{rs} \pdv{E(\bmf, \bmkp^*, \bmth^*)}{\theta_k}{\kappa_{rs}} \pdv{\kappa_{rs}^*(\bmf)}{F_d} + 
 \sum_l \pdv{E(\bmf, \bmkp^*, \bmth^*)}{\theta_k}{\theta_l} \pdv{\theta_l^*(\bmf)}{F_d}= - \pdv{E(\bmf, \bmkp^*, \bmth^*)}{\theta_k}{F_d}.
 \end{split}
\end{equation}
The solution of this equation is
\begin{align}
\label{eq: solution of response}
 \left( \pdv{\bmkp^*(\bmf)}{F_d}, \pdv{\bmth^*(\bmf)}{F_d}
 \right)^T
 = 
 \left(
 \begin{array}{c|c}
  \mathbb{E}_{\bmkp, \bmkp} &  \mathbb{E}_{\bmkp, \bmth} \\ \hline
  \mathbb{E}_{\bmth, \bmkp} &
  \mathbb{E}_{\bmth, \bmth}
 \end{array}
 \right)^{-1}
 \left(
 \mathbb{V}_{\bmkp}^{(d)}, \mathbb{V}_{\bmth}^{(d)}
 \right)^T,
\end{align}
where the submatrices are defined as
\begin{equation}
 \begin{split}
  \left(\mathbb{E}_{\bmkp, \bmkp}\right)_{pq,rs} \equiv \pdv{E(\bmf, \bmkp^*, \bmth^*)}{\kappa_{pq}}{\kappa_{rs}},
 \left(\mathbb{E}_{\bmkp, \bmth}\right)_{pq, l} \equiv \pdv{E(\bmf, \bmkp^*, \bmth^*)}{\kappa_{pq}}{\theta_l}, \\
 \left(\mathbb{E}_{\bmth, \bmkp}\right)_{k,rs} \equiv \pdv{E(\bmf, \bmkp^*, \bmth^*)}{\theta_k}{\kappa_{rs}},
 \left(\mathbb{E}_{\bmth, \bmth}\right)_{kl} \equiv \pdv{E(\bmf, \bmkp^*, \bmth^*)}{\theta_k}{\theta_l}
 \end{split}
\end{equation}
for all MO pairs $(p,q)$ and $k,l=1,\cdots,M$,
and the vectors $\mathbb{V}_{\bmkp}^{(d)}, \mathbb{V}_{\bmth}^{(d)}$ are defined as
\begin{equation}
 \begin{split}
\left( \mathbb{V}_{\bmkp}^{(d)} \right)_{pq} \equiv - \pdv{E(\bmf, \bmkp^*, \bmth^*)}{\kappa_{pq}}{F_d},
\left( \mathbb{V}_{\bmth}^{(d)} \right)_k \equiv - \pdv{E(\bmf, \bmkp^*, \bmth^*)}{\theta_k}{F_d}.
 \end{split}
\end{equation}
As we reviewed in the previous section, all quantities in the right hand side of Eq.~\eqref{eq: solution of response} can be evaluated by the standard technique for evaluating expectation values on quantum computers without any ancillary qubit.
Therefore, we can calculate the responses of the optimal parameters [Eq.~\eqref{eq: response of optimal params}] with classical computers by putting the evaluated values in the right hand side of Eq.~\eqref{eq: solution of response}.
Note that the inverse matrix in Eq.~\eqref{eq: solution of response} can be numerically problematic when the value of each entry of the matrix has noise.
Error analysis including this point will be presented in Sec.~\ref{subsubsec: error theory analytical}.

Finally, combining the evaluated values in Eq.~\eqref{eq: evs in 2nd deriv.} and the computed values of $\pdv{\bmkp^*}{F_d}, \pdv{\bmth^*}{F_d}$ in Eq.~\eqref{eq: formula 2nd deriv.} gives the second-order derivative of the OO-VQE energy.

In summary, the calculation of the analytical second-order derivative of the OO-VQE energy proceeds as follows:
\begin{enumerate}
 \item Evaluate the partial derivatives listed in~\eqref{eq: evs in 2nd deriv.} by quantum computers.
 \item Evaluate the right hand side of Eq.~\eqref{eq: solution of response} by quantum computers, and compute the responses of the optimal parameters (the left hand side) by classical computers.
 \item Put the values of those responses into Eq.~\eqref{eq: formula 2nd deriv.} and compute the second-order derivative of the energy by classical computers.
\end{enumerate}
We note that our derivation of the analytical second-order derivative of OO-VQE energy is essentially parallel to that of MCSCF, a classical counterpart of OO-VQE~\cite{Helgaker1989, Szalay2012}.
Our theoretical contribution in this study is to present the explicit formula of the second-order derivative of the OO-VQE energy and the way to evaluate the necessary quantities with classical and quantum computers.

%%%%% ----- %%%%%
\section{Application to \pol of molecules \label{sec: pol application}}
In this section, we consider the polarizability of molecules. 
The \pol is the second-order derivative of the energy with respect to the external electric field applied to the molecule.
We first derive an analytical formula of the \pol for OO-VQE by using the formula explained in the previous section.
We then perform numerical calculation of the \pol based on classical simulation of quantum circuits by taking the water molecule as an example.
We also conduct numerical simulation for the \pol of thiophene \ce{C4H4S} and furan \ce{C4H4O} molecules.
The \pol calculated by the numerical simulation is converted to the refractive indices of the molecules, and the results exhibit the same tendency as experimental observations.

\subsection{Analytical formulation of \pol}
The molecular Hamiltonian under the application of the electric field $\bmf = (F_x, F_y, F_z)$ is
\begin{equation}
\label{eq: ham for pol}
 \hat{H}_\mr{full}^{(E)}(\bmf, \bmkp) \equiv \hat{H}_\mr{full,0}^{(E)}(\bmkp) - \sum_{d=x,y,z} F_d \hat{\mu}_d(\bmkp),
\end{equation}
where
\begin{align}
 \hat{H}_{\mr{full},0}^{(E)}(\bmkp) &\equiv E_\mr{c} + \sum_{p,q=1}^{N_\mathrm{MO}} \sum_{\sigma=\uparrow, \downarrow} h_{pq}(\bmkp)\hat{a}_{p\sigma}^\dag\hat{a}_{q\sigma}
 + \frac{1}{2} \sum_{p,q,r,s=1}^{N_\mathrm{MO}} \sum_{\sigma,\tau=\uparrow, \downarrow} g_{pqrs}(\bmkp)\hat{a}^\dag_{p\sigma}\hat{a}^\dag_{q\tau}\hat{a}_{r\tau}\hat{a}_{s\sigma},
\end{align}
is the Hamiltonian without electric field and
\begin{equation}
 \hat{\mu}_d = \sum_{pq, \sigma} M_{pq}^{(d)}(\bmkp) \hat{a}_{p\sigma}^\dag\hat{a}_{q\sigma} 
\end{equation}
is the dipole moment operator in the direction $d=x,y,z$.

$E_\mr{c}, h_{pq}(\bmkp), g_{pqrs}(\bmkp)$ and $M_{pq}^{(d)}(\bmkp)$ are defined by
\begin{align*}
E_\mr{c} &= \sum_{I,J} \frac{Z_I Z_J}{|\bm{R}_I - \bm{R}_J|}, \\
h_{pq}(\bmkp) &= \int d\bm{r} \phi_p^*(\bm{r}; \bmkp) \left( -\frac{\nabla^2_{\bm{r}}}{2} + \sum_I \frac{Z_I}{|\bm{r} - \bm{R}_I|} \right)  \phi_q(\bm{r}; \bmkp), \\
 g_{pqrs}(\bmkp) & = \int d\bm{r}_1 d\bm{r}_2 \phi_p^*(\bm{r}_1; \bmkp) \phi_q^*(\bm{r}_2; \bmkp) \frac{1}{|\bm{r}_1 - \bm{r}_2|} \phi_r(\bm{r}_2; \bmkp) \phi_s(\bm{r}_1; \bmkp), \\
 M_{pq}^{(d)}(\bmkp) &= \int d\bm{r} \, \bm{r}_d \, \phi_p^*(\bm{r}; \bmkp) \phi_q(\bm{r}; \bmkp),
\end{align*}
where $\bm{R}_I \,(Z_I)$ are coordinates (charge) of the $I$-th nuclei in the molecule, $\phi_p(\bm{r}_1; \bmkp)$ is the molecular orbital $p$ depending on the orbital parameters $\bmkp$, and $\bm{r}_d$ is $d$-coordinate of $\bm{r}$ for $d=x,y,z$.
These values can be efficiently calculated by classical computers.
We note that we fix the positions of the nuclei of the molecule and assume the Born-Oppenheimer approximation.

The \pol is defined by the second-order derivative of energy,
\begin{equation}
\label{eq: def of pol}
 P_{dd'} \equiv - \pdv{E^*(\bmf)}{F_d}{F_{d'}},
\end{equation}
for $d,d'=x,y,z$.
According to the definition Eq.~\eqref{eq: ham for pol}, we see
\begin{align}
\label{eq: evs for pol}
  \pdv{E(\bmf, \bmkp^*, \bmth^*)}{F_d}{F_{d'}} = 0,\;
  \pdv{E(\bmf, \bmkp^*, \bmth^*)}{\kappa_{pq}}{F_d} = - \pdv{\mu_d(\bmkp^*,\bmth^*) }{\kappa_{pq}},\;
  \pdv{E(\bmf, \bmkp^*, \bmth^*)}{\theta_k}{F_d} = - \pdv{\mu_d(\bmkp^*,\bmth^*)}{\theta_k},
\end{align}
where $\mu_d(\bmkp^*,\bmth^*) \equiv \ev{\hat{\mu}_d(\bmkp^*(\bmf))}{\Psi}, \ket{\Psi}= \ket{\mr{vac}}_\mr{vir} \otimes \ket{\psi(\bmth^*(\bmf))} \otimes\ket{\uparrow\downarrow}_\mr{core}$
is the expectation value of the dipole moment operator at the optimal parameters.
Substituting Eqs.~\eqref{eq: def of pol} \eqref{eq: evs for pol} into Eqs~\eqref{eq: formula 2nd deriv.} and \eqref{eq: solution of response} gives the analytical formula for the polarizability:
\begin{align}
\label{eq: formula pol}
 P_{dd'} =
 \left(
 \pdv{\mu_d(\bmkp^*, \bmth^*)}{\bmkp},  \pdv{\mu_d(\bmkp^*, \bmth^*)}{\bmth}
 \right)
 \left(
 \begin{array}{c|c}
  \mathbb{E}_{\bmkp, \bmkp} &  \mathbb{E}_{\bmkp, \bmth} \\ \hline
  \mathbb{E}_{\bmth, \bmkp} &
  \mathbb{E}_{\bmth, \bmth}
 \end{array}
 \right)^{-1}
 \left(
 \begin{array}{c}
\pdv{\mu_{d'}(\bmkp^*, \bmth^*)}{\bmkp} \\  \pdv{\mu_{d'}(\bmkp^*, \bmth^*)}{\bmth}
 \end{array}
 \right).
\end{align}
This simple equation is one of our main results.

\subsection{Numerical calculations for the \pol and refractive index of molecule}
We present results of numerical simulations for calculating the polarizability.
We first numerically calculate the \pol of the water molecule by simulating quantum circuits and their outputs during OO-VQE with classical computers.
The \pol computed by our formula agrees well with the reference value computed by MCSCF (the classical counterpart of OO-VQE).
Next, as a first step towards potential industrial applications in the future, we perform numerical calculation of the \pol of thiophene \ce{C4H4S} and furan \ce{C4H4O} molecule, which can form polymers with several industrial interests.
Moreover, we estimate the refractive indices of the two molecules based on the calculated \pol and the experimental values of the molecular density.
The result for the refractive indices qualitatively matches with experimental observations.

\subsubsection{Checking validity of the formula with water molecule}
\begin{table*}[ht]
\caption{The numerical result of the energy $E$, the dipole moment $\mu = \sqrt{\mu_x^2 + \mu_y^2 + \mu_z^2}$, and the \pol of the water molecule. \label{tab: H2O result} }
\begin{tabular}{cccccc} 
\hline  \hline \\
Method & $E$ [Ha] & $\mu$ (a.u.) & $P_{xx}$ (a.u.) & $P_{yy}$ (a.u.)& $P_{zz}$ (a.u.)
\\ \hline % [-1.8ex] 
Quantum (OO-VQE) & -75.99956 & 0.97083 & 1.41686 & 6.59715 & 3.89083 \\ 
Classical (MCSCF) & -75.99956 & 0.97084 & 1.41686 & 6.59714 & 3.89082  \\ 
\hline  
\end{tabular} 
\end{table*} 

We took the water molecule $\ce{H2O}$ as an example to numerically check the validity of our formula for the polarizability.
First, we performed numerical simulation of OO-VQE for electronic states of the water molecule. 
The geometry of the molecule was taken as the most stable one at the level of Hartree-Fock/6-31G, retrieved from the CCCBDB database~\cite{cccbdb} (coordinates are given in Supporting Information).
We employed the 6-31G basis set and took the active space consisting of three orbitals (HOMO-1 (\ce{a1} symmetry), HOMO (\ce{b1}), and LUMO (\ce{a1})) and four electrons, which is the minimal choice around HOMO that includes more than two orbitals with the same symmetry.
The orbital rotation and construction of the molecular Hamiltonian were performed by the numerical package PySCF~\cite{Sun2018_pyscf}.
We used unitary coupled-cluster singles and doubles ansatz~\cite{peruzzo2014variational,anand2022} with the first-order Trotterization as a trial state $\ket{\psi(\bmth)}$, and there were five circuit parameters in total (see Supporting Information for details).
Jordan-Wigner transformation~\cite{Jordan1928} was used for mapping the fermion Hamiltonian to the qubit one.
The optimization of the circuit parameters $\bmth$ was done with the BFGS (Broyden–Fletcher–Goldfarb–Shanno) method implemented in SciPy~\cite{SciPy2020}.
All outputs of the quantum circuits for executing OO-VQE and calculating the \pol in this section were simulated by classical computers using the numerical package Qulacs~\cite{Suzuki2021qulacsfast}, assuming no noise and statistical fluctuations in the outputs.
We note that the matrix $\mathbb{E}$ in Eq.~\eqref{eq: formula pol} became singular (not full-rank) in our simulation.
We took the pseudo-inverse of $\mathbb{E}$ for such a case, i.e., we took the inverse of the matrix by ignoring the singular values of $\mathbb{E}$ smaller than the threshold $r_\mr{cond} \cdot \sigma_\mr{max}$, where $\sigma_\mr{max}$ is the largest singular value of $\mathbb{E}$.
We set $r_\mr{cond} = 10^{-5}$ in the numerical simulation in this section.

The results of OO-VQE and the value of the \pol calculated by Eq.~\eqref{eq: formula pol} are summarized in Table~\ref{tab: H2O result}.
All of the results agree well with the reference values of MCSCF calculated by DALTON~\cite{daltonpaper} Release v2020.0, a classical computational package for quantum chemistry.
These results validate our analytical formula for the polarizability.

\subsubsection{Polarizability and refractive indices of thiophene and furan}
\begin{table*}
\caption{The numerical result of the energy $E$, the dipole moment $\mu = \sqrt{\mu_x^2 + \mu_y^2 + \mu_z^2}$, the polarizability, and the refractive index $n$ of the thiophene \ce{C4H4S}  and furan \ce{C4H4O}. \label{tab: C4H4 result} }
\begin{tabular}{c|ccccccc} 
\hline  \hline
Molecule & Method & $E$ [Ha] & $\mu$ (a.u.) & $P_{xx}$ (a.u.) & $P_{yy}$ (a.u.)& $P_{zz}$ (a.u.) & 
$n$ \\ \hline 
\multirow{2}{*}{\ce{C4H4S}}
& Quantum (OO-VQE) & -551.24415 & 0.67117 & 20.74094 & 53.11627 & 60.33725 & 1.344 \\ 
& Classical (MCSCF) & -551.24440 & 0.67228 & 20.74147 & 53.12157 & 60.25941 & 1.343 \\ \hline
\multirow{2}{*}{\ce{C4H4O}}
& Quantum (OO-VQE) & -228.57737 & 0.62253 & 14.60875 & 40.95479 & 41.00646 & 1.267 \\ 
& Classical (MCSCF) & -228.57761 & 0.62404 & 14.60801 & 40.95588 & 40.94519 & 1.267
\end{tabular} 
\end{table*} 

We investigated the \pol and refractive indices of thiophene \ce{C4H4S} and furan \ce{C4H4O}, both of which constitute typical monomers of transparent polymer materials.
Although the system size we study here is small enough to be easily handled by classical computers, the calculation of the refractive indices for two molecules can be viewed as a steady but important step for designing transparent materials and utilizing quantum computers in industry.
The geometries of these molecules were again taken as the most stable ones at the level of Hartree-Fock/6-31G, retrieved by the CCCBDB database~\cite{cccbdb} (the three-dimensional coordinates are given in Supporting Information).
We employed the 6-31G basis set and chose the active space of four orbitals consisting of HOMO-1, HOMO, LUMO, and LUMO+1 with four electrons, which is the minimal choice that includes the orbitals with the same symmetries as HOMO (\ce{a2)} and LUMO (\ce{b1})~\cite{Telesca2001}.
The other conditions for the numerical simulation were the same as those for the water molecule.

The results are summarized in Table~\ref{tab: C4H4 result}.
The energy, dipole moment, and the \pol calculated by OO-VQE with our formula agree with the reference values computed by MCSCF.
The deviations between OO-VQE and MCSCF are within 0.3\%.
The reason for these deviations is because the active space solver of OO-VQE, namely VQE with the unitary coupled cluster singles and doubles ansatz with the first-order Trotterization, is not as precise as that of MCSCF (i.e., full configuration interaction) in this case.
Finally, the reflective index $n$ was estimated by assuming the Lorentz-Lorenz formula~\cite{kitaev1995quantum},
\begin{equation}
 n = \sqrt{ \frac{1+2\phi}{1-\phi} }, \quad \phi = \frac{4\pi}{3} N_m \alpha,
\end{equation}
where $N_m$ is the number of molecules per volume and $\alpha = (P_{xx} + P_{yy} + P_{zz})/3$ is the mean \pol in CGS units (expressed in volume).
The value of $N_m$ was computed by using the experimental molecular density, \ce{1.06494} g/cm$^3$ for thiophene~\cite{crc76th} and \ce{0.9514} g/cm$^3$ for furan~\cite{crc88th}.
The results are shown in the most right column of Table~\ref{tab: C4H4 result}.
These values are consistent with the experimental observation that the thiophene molecule has a larger refractive index~(1.52684 for thiophene~\cite{merck1996} and 1.4214 for furan~\cite{industrial1985}).

%%%%% ----- %%%%%
\section{Analysis of error in analytical derivative and numerical derivative \label{sec: error analysis}}
In this section, we analyze errors in estimated values of the second-order derivative of the OO-VQE energy for two methods; one is the analytical derivative we proposed in this study and the other is the numerical derivative, or brute-force numerical differentiation of the energy with small finite difference.
We first theoretically investigate the errors in the analytical and numerical derivatives.
The different scaling exponents of the error in the \pol may suggest the advantage of the analytical derivative over the numerical one.
Next, we perform numerical experiments comparing the analytical and numerical derivatives by taking \ce{LiH} molecule as an example.
The numerical data indicate that the analytical derivative requires fewer measurements (quantum circuit execution) compared with the numerical derivative to achieve the same fixed accuracy for the estimated polarizability.

\subsection{Theoretical analysis of error scaling}
Expectation values evaluated by quantum computers inevitably fluctuate because of the sampling error.
Even more, currently-available quantum computers have various noise sources that make the expectation values deviate from the exact one.
Here we provide the (noise) error analysis of the second-order derivative of OO-VQE energy.
In the following, we write an estimated value of a quantity $Q$ as $\tilde{Q}$.

\subsubsection{Analytical derivative \label{subsubsec: error theory analytical}}
We introduce the notation for unifying the orbital and circuit parameters,
$\bm{X} = (X_1, \cdots, X_{N_p}) \equiv (\bmkp, \bmth)$.
We assume that there is no error in the optimal parameters $\bm{X}^* = (\bmkp^*, \bmth^*)$ and that estimation errors of $\pdv{\mu(\bmkp^*, \bmth^*)}{X_a}$ and $\pdv{E(\bmf, \bmkp^*, \bmth^*)}{X_a}{X_b}$ are bounded as
\begin{equation}
 \abs{\wt{\pdv{\mu}{X_a}} - \pdv{\mu}{X_a}} \leq \epsilon_{\partial\mu}, \quad
  \abs{\wt{\pdv{E}{X_a}{X_b}} - \pdv{E}{X_a}{X_b}} \leq \epsilon_{\mr{Hesse}}, 
\end{equation}
for all $a,b = 1, ..., N_p$.
As described in Eq.~(C7) of Ref.~\citenum{mitarai2020theory} or Section 5.8 of Ref.~\citenum{horn2012matrix},
when $\bm{b} =A \bm{y}$ and $\bm{b}+ \Delta\bm{b}  = (A+\Delta A) \bm{y}'$, the error of $\bm{y}$ is written as
\begin{equation}
 \frac{\abs{\bm{y}' - \bm{y}}_2}{|\bm{y}|_2} \leq 
 \frac{\kappa(A)}{1-\kappa(A)\frac{|\Delta A|_F}{|A|_F}} \left( \frac{|\Delta\bm{b}|_2}{|\bm{b}|_2} + \frac{|\Delta A|_F}{|A|_F} \right),
\end{equation}
where $\kappa(A)$ is a condition number of the matrix $A$ defined as $\kappa(A) = |A^{-1}|_F |A|_F$, the norm for matrices is Frobenius norm $|A|_F = \sqrt{\Tr(A^\dag A)} = \sqrt{\sum_{ab} |A_{ab}|^2}$, and the norm for vector is Euclidean $l^2$-norm $|\bm{y}|_2 = \sqrt{\sum_a |y_a|^2}$.
\footnote{This inequality holds for all matrix norms $|...|_\mr{mat}$ and their consistent vector norms $|...|_\mr{vec}$ satisfying $\abs{A\bm{y}}_\mr{vec} \leq |A|_\mr{mat} |\bm{y}|_\mr{vec}$.
Here we use Frobenius norm and $l^2$-norm because they are easy to evaluate.}
In the case of the \pol $P_{dd'}$ [Eq.~\eqref{eq: solution of response}], we see
\begin{equation}
\bm{b} = - \pdv{E(\bmf, \bmkp^*, \bmth^*)}{F_d}{\bm{X}} = \pdv{\mu_d(\bmkp^*, \bmth^*)}{\bm{X}}, A =\pdv{E(\bmf, \bmkp^*,\; \bmth^*)}{\bm{X}}{\bm{X}},\;
\bm{y} = \pdv{\bm{X}^*(\bmf)}{F_d}.
\end{equation}
So the estimation error of $\pdv{\bm{X}^*(\bmf)}{F_d}$ is bounded as
\begin{equation}
 \abs{ \wt{\pdv{\bm{X}^*(\bmf)}{F_d}} - \pdv{\bm{X}^*(\bmf)}{F_d}}_2 \leq
 \abs{\pdv{\bm{X}^*(\bmf)}{F_d}}_2
 \frac{\kappa(A)}{1-\kappa(A) \frac{N_p \epsilon_\mr{Hesse}}{|A|_F}}
 \left( \frac{ \sqrt{N_p} \epsilon_{\partial\mu }}{\abs{\pdv{\mu_i}{\bm{X}}}_2}
 + \frac{N_p \epsilon_\mr{Hesse}}{|A|_F} \right).
\end{equation}
Since Eq.~\eqref{eq: formula pol} can be read as
\begin{equation}
 P_{dd'} = \pdv{\bm{X}^*(\bmf)}{F_d} \cdot \pdv{\mu_{d'}(\bmkp^*, \bmth^*)}{\bm{X}},
\end{equation}
the estimation error of the \pol is
\begin{align}
\label{eq: error of pol}
\begin{split}
 \abs{\wt{P_{d{d'}}} -P_{d{d'}}} \leq &
 \abs{ \wt{\pdv{\bm{X}^*(\bmf)}{F_d}} - \pdv{\bm{X}^*(\bmf)}{F_d}}_2 \cdot \abs{\pdv{\mu_{d'}}{\bm{X}}}_2
 + \abs{\pdv{\bm{X}^*(\bmf)}{F_d}}_2 \cdot \abs{ \wt{\pdv{\mu_{d'}}{\bm{X}}} - \pdv{\mu_{d'}}{\bm{X}} }_2 \\
 & + \abs{ \wt{\pdv{\bm{X}^*(\bmf)}{F_d}} - \pdv{\bm{X}^*(\bmf)}{F_d}}_2 \cdot \abs{ \wt{\pdv{\mu_{d'}}{\bm{X}}} - \pdv{\mu_{d'}}{\bm{X}} }_2 \\
 \leq &
 \abs{\pdv{\bm{X}^*(\bmf)}{F_d}}_2
 \frac{\kappa(A)}{1-\kappa(A) \frac{N_p \epsilon_\mr{Hesse}}{|A|_F}}
 \left( \frac{ \sqrt{N_p} \epsilon_{\partial\mu }}{\abs{\pdv{\mu_i}{\bm{X}}}_2}
 + \frac{N_p \epsilon_\mr{Hesse}}{|A|_F} \right)
 \left( \abs{\pdv{\mu_{d'}}{\bm{X}}}_2 + \sqrt{N_p} \epsilon_{\partial\mu} \right) \\
 & + \abs{\pdv{\bm{X}^*(\bmf)}{F_d}}_2 \cdot \sqrt{N_p} \epsilon_{\partial\mu}.
\end{split}
\end{align}
We can observe from this equation that the error of the \pol can be made small when the denominator of the first term is not so small, i.e.,
\begin{equation*}
 1-\kappa(A) \frac{N_p \epsilon_\mr{Hesse}}{|A|_F} \sim 1 \: \Rightarrow \:
 \epsilon_\mr{Hesse}  \ll \frac{|A|_F}{N_p \kappa(A)}.
\end{equation*}
This is because the other terms are proportional to the errors of the estimated quantities on quantum computers, $\epsilon_{\partial\mu }$ and $\epsilon_\mr{Hesse}$, which can be made small in principle (e.g., by increasing the number of measurements).
Although the condition number of $A$ (the energy Hesse matrix) is not known \textit{a priori}, the above equation gives a necessary condition for the suitable value of $\epsilon_\mr{Hesse}$ to suppress the error of the polarizability.

\subsubsection{Numerical derivative}
Apart from using the analytical derivative, it is possible to consider the numerical derivative with a small finite difference to calculate the derivative of energy.
Let us focus on the \pol for $\bmf = \bm{0}$.
To calculate $P_{dd'}$ with the numerical derivative method, one performs OO-VQE twice for two different finite fields $F_d = \pm h, F_{k \neq d} = 0$ and calculates the difference of the expectation value $\mu_{d'}$:
\begin{equation} \label{eq: num. dev. of mu}
 P_{dd'} = - \pdv{\mu_{d'}(\bmkp^*, \bmth^*)}{F_d} =
 - \frac{\mu_{d'}^*(F_d = h) - \mu_{d'}^*(F_d = -h)}{2h} 
 + \order{h^2}
\end{equation}
where $\mu_{d'}^*(\bmf) \equiv \mu_{d'}(\bmkp^*(\bmf), \bmth^*(\bmf))$ is the expectation value of the dipole moment operator.
Taylor's theorem leads to 
\begin{equation} \label{eq: num deriv rough bound}
 \abs{\wt{P_{d{d'}}} - P_{d{d'}} } \leq
 \frac{\epsilon_\mu}{h} + \frac{h^2}{6} \max_{F_d \in [-h, h], F_{k\neq d}=0} \abs{\pdv[3]{\mu_{d'}^*(\bmf)}{F_d}},
\end{equation}
where we define $\epsilon_\mu$ by  $|\widetilde{\mu_{d'}} - \mu_{d'}| \leq \epsilon_{\mu}$.
To bound the estimation error of $P_{d{d'}}$ within $\epsilon_\mr{Pol}$, i.e., $ \abs{\wt{P_{d{d'}}} - P_{d{d'}} } \leq \epsilon_\mr{Pol}$, we need
\begin{equation}
\label{eq: ep_mu}
 \epsilon_\mu = h \qty( \epsilon_\mr{Pol} - \frac{h^2}{6} \max_{F_d \in [-h, h], F_{k\neq d}=0} \abs{\pdv[3]{\mu_{d'}^*(\bmf)}{F_d}} ),
\end{equation}
with the assumption that the right hand side is positive.
We further assume that
\begin{equation}
\max_{F_d \in [-h, h], F_{k\neq d}=0} \abs{\pdv[3]{\mu_{d'}^*(\bmf)}{F_d}} = C_3
\end{equation}
can be considered constant with respect to $h$ in the region of interest.
In this case, the value of $h$ which maximizes the right hand side of Eq.~\eqref{eq: ep_mu} is
$h = \sqrt{2\epsilon_\mr{Pol}/ C_3}$,
and the value of $\epsilon_\mu$ for that $h$ is
\begin{equation}
 \label{eq: ep_mu_best}
 \epsilon_\mu^* = \frac{2\sqrt{2}}{3}
 \frac{\epsilon_{\mr{Pol}}^{3/2}}{C_3^{1/2}} 
 \quad \Leftrightarrow \quad
 \epsilon_\mr{Pol} = \frac{\sqrt[3]{9C_3}}{2} {\epsilon_\mu^*}^{2/3}.
\end{equation}

Comparing the scaling of error in the analytical and numerical derivatives, therefore, the error of the \pol in the numerical derivative scales with $\epsilon_\mu^{2/3}$ while that in the analytical derivative [Eq.~\eqref{eq: error of pol}] scales with $\epsilon_{\partial \mu}$ and $\epsilon_\mr{Hesse}$.
The exponent of the power $2/3$ for the numerical derivative may mean that the numerical derivative is more vulnerable to the noise and fluctuation of expectation values evaluated on quantum computers although many prefactors appearing in the formulas \eqref{eq: error of pol} and \eqref{eq: ep_mu_best} can affect the conclusion.
We note that this analysis is similar to those of Appendix C.2 of Ref.~\citenum{mitarai2020theory} and Appendix C of Ref.~\citenum{tamiya2021}.

\subsection{Numerical experiment for LiH molecule}
\begin{figure}[t]
\centering
\includegraphics[width=0.5\textwidth]{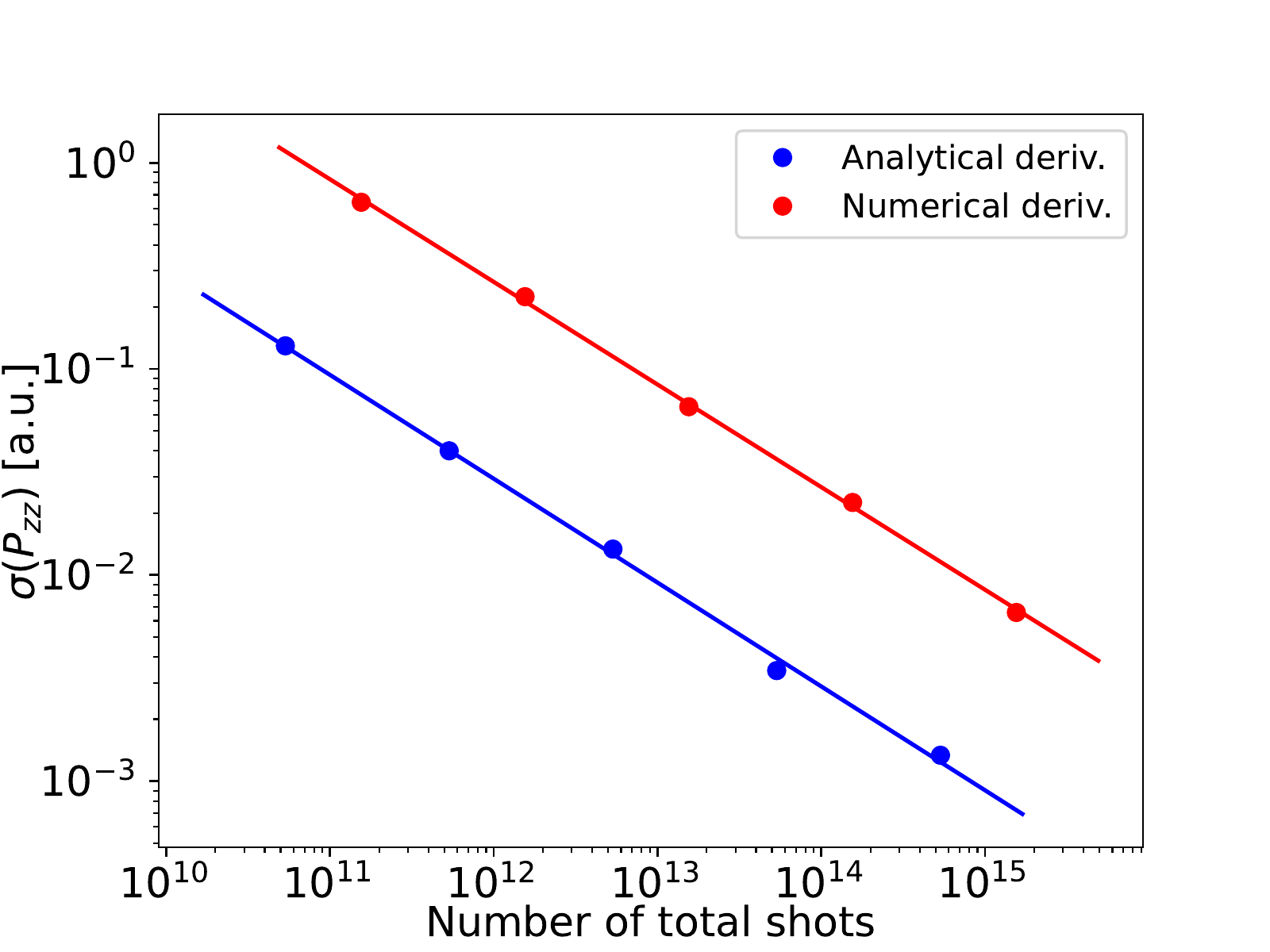}
\caption{
The sample standard deviation of the \pol $P_{zz}$ of LiH molecule versus the total number of shots required to evaluate $P_{zz}$. Lines are the results of the linear regression on the log-log plot.
\label{fig: std Pzz}
}
\end{figure}

When evaluating expectation values of observables on quantum computers (especially NISQ devices), we repeatedly prepare a quantum state and execute specific quantum circuits to estimate the expectation values.
Each run of the quantum circuit is called a (measurement) shot, and it is important to reduce the number of shots to save computational time.
Here, we compare the required number of shots in quantum computers to calculate the \pol with the same fixed precision by using the analytical and numerical derivatives.
We did this by numerical simulation for \ce{LiH} molecule.

In the numerical experiment, we considered \ce{LiH} molecule with the 6-31G basis set.
The bond distance was taken as 3.013924 Bohr (the most stable structure at the level of Hartree-Fock/6-31G) and the molecule was aligned in $z$-direction.
The active space was set to three orbitals (HOMO-1, HOMO, and LUMO) with four electrons, which includes 1s orbital of \ce{Li} and the bonding and anti-bonding orbitals of \ce{LiH}.
To perform OO-VQE, the unitary coupled-cluster singles and doubles ansatz~\cite{peruzzo2014variational,anand2022} with the first-order Trotterization was again employed as a trial state $\ket{\psi(\bmth)}$, and the total number of the circuit parameters was five.
In both cases of the analytical derivative and the numerical derivative, OO-VQE was simulated by classical computers with assuming no noise and error in quantum circuits as in Sec.~\ref{sec: pol application}, and the optimal parameters $\bmkp^*$ and $\bmth^*$ were determined.
We do not consider the number of shots to perform OO-VQE itself and focus on the number of shots to calculate the $zz$-component of the \pol $P_{zz}$ for given (exact) orbital and circuit parameters $\bmkp^*$ and $\bmth^*$.

After the performance of OO-VQE without any noise, we simulated the calculation of the \pol with including the effect of the fluctuation of outputs of quantum circuits.
We assumed that the same number of the shots, $n_\text{1Pauli}$, was consumed for estimating expectation values of all distinct Pauli operators included in the observable.
The expectation values of the observables in the formulas of the analytical and numerical derivatives were estimated by simulating the measurement results of quantum circuits under this assumption, resulting in the fluctuation (error) of the estimated expectation values.
We note that no other noise such as the depolarizing noise was included in the numerical simulation.
Further details are described in Supporting Information.

The number of total shots to obtain $P_{zz}$ was estimated as follows.
For the analytical derivative, we count the number of shots to measure all quantities in the right hand side of Eq.~\eqref{eq: formula pol}.
Namely, all independent elements of the 1,2-RDMs [Eq.~\eqref{eq: def of RDMs}] of the state $\ket{\psi(\bmth^*)}$ were measured by considering the exchange symmetry among indices.
The first-order and second-order $\theta$ derivatives of the 1,2-RDMs were also measured.
As described in the previous section, we set a threshold $r_\mr{cond}$ for the singular values of $\mathbb{E}$ when taking the (pseudo-)inverse of $\mathbb{E}$ in the analytical derivative.
We chose $r_\mr{cond}=10^{-3}$ in this section.
For the numerical derivative, we count the number of shots to measure all independent elements of the 1,2-RDM for the optimal states at $F_z = \pm h$.
The finite difference $h$ was chosen as $h=0.001$ (in atomic unit) so that the deviation of the \pol compared to the exact one (obtained by MCSCF) due to the finite difference becomes smaller than 0.01 in atomic units.
We note that the terms in the right hand side of Eq.~\eqref{eq: num. dev. of mu} can be measured with only 1-RDM, but running OO-VQE requires both 1-RDM and 2-RDM in the course of it, so we include the number of shots to measure 2-RDM even for the numerical derivative.

We simulated the calculation of the \pol $P_{zz}$ for 100 times and computed the sample standard deviation, $\sigma(P_{zz})$, with varying the number of shots for a single Pauli operator, $n_\mr{1Pauli}$.
We chose $n_\mr{1Pauli} = 10^5, 10^6, 10^7, 10^8, 10^9$ for the analytical derivative and $n_\mr{1Pauli} = 10^8, 10^9, 10^{10}, 10^{11}, 10^{12}$ for the numerical derivative.
For each method, the total number of shots was calculated by accumulating the number of shots for evaluating the Pauli operators in the corresponding formula.
The result is shown in Fig.~\ref{fig: std Pzz}.
The analytical derivative requires fewer measurement shots to reach the same fixed precision (standard deviation) of $P_{zz}$ than the numerical derivative does.
This possibly indicates the advantage of the analytical derivative over the numerical derivative.

Moreover, the standard deviations of $P_{zz}$ for both derivatives exhibit the power-law decay with the total number of shots, $n_\text{total shots}$.
Namely, it seems to hold $\sigma(P_{zz}) \propto (n_{\text{total shots}})^{-\alpha}$.
The least squares fitting on the log-log plot in Fig.~\ref{fig: std Pzz} indicates that the exponent $\alpha$ for the analytical derivative is $\alpha \sim 0.504(13)$ and that for the numerical derivative is $\alpha \sim 0.498(7)$.
We can expect that the estimation errors of the observables in Eqs.~\eqref{eq: error of pol} and \eqref{eq: num deriv rough bound} obey
\begin{equation}
 \epsilon_{\mu}, \epsilon_{\partial \mu}, \epsilon_\mr{Hesse} \propto (n_{\text{total shots}})^{-1/2},
\end{equation}
because the standard deviation of the expectation value of the single Pauli operator decays as $(n_\mr{1Pauli})^{-1/2}$. 
The exponents of the power-law decay for both the analytical and numerical derivatives in the numerical simulations are therefore consistent with our analysis [Eqs.~\eqref{eq: error of pol} and \eqref{eq: num deriv rough bound}] (note that the size of the finite difference $h$ was fixed in our experiment).

Finally, we comment on possible improvements to reduce the seemingly large number of shots ($\sim 10^{15}$) in Fig.~\ref{fig: std Pzz}, although the purpose of the plot is a comparison between the analytical and numerical derivatives.
There are several ways to reduce the actual runtime of quantum computers in our algorithm.
First, the ``shots" counted here can be executed mostly in parallel because the evaluation of each component of the matrix and vectors in Eq.~\eqref{eq: formula pol} is independent of the others.
The actual runtime will be greatly reduced when there are multiple quantum devices.
Second, we are able to employ various techniques to reduce the number of shots to evaluate the expectation values of the observables, which can improve the number of shots by several orders of magnitude~\cite{Kohda2022}.

%%%%% ----- %%%%%
\section{Discussion, summary and outlook \label{sec: summary}}
Before ending the article, let us discuss the relationship of our results to several previous studies.
In Ref.~\citenum{mitarai2020theory}, the theory of the analytical derivative of the energy obtained by usual VQE, including the second-order one, was developed.
It did not consider the orbital optimization, while our result explicitly treats the orbital optimization and the orbital parameters. 
We also note that Ref.~\citenum{Mizukami2020}, which is one of the papers proposing OO-VQE, mentioned the first-order derivative of the energy.
As for the calculation of the polarizability, O'Brien {\it et~al.}~\cite{o2019calculating} calculated the \pol of the hydrogen molecule by using the technique called the sum-over-states, which requires a lot of excited states of the Hamiltonian (exponentially large number of eigenstates, in principle).
Another study for calculating the \pol is Huang {\it et~al.}~\cite{huang2022simulating}, where the authors calculated the dynamical \pol by preparing a quantum state proportional to the perturbed ground state by the dipole operator, $\hat{\mu}_d \ket{\psi_\text{ground state}}$ with the variational method. 
Compared with it, our approach has an advantage in that we perform the variational optimization only once (for finding the ground state).
It will reduce the cost of quantum computation.

In summary, we developed a quantum-classical hybrid algorithm to calculate the second-order derivative of the energy obtained by OO-VQE.
The analytical formula for the derivative and the procedures to evaluate all terms in the formula on quantum computers were explained. 
We applied the formula of the analytical derivative to the polarizability and validated the formula for the \pol by numerical simulations.
Moreover, the \pol and refractive indices of thiophene and furan molecules, both of which are of potential industrial interest, were calculated by numerical simulations using our formula.
Finally, we analyzed the effect of the error on the estimated \pol by our analytical derivative and the numerical derivative with a finite difference.
The analytical derivative proposed by us is advantageous over the numerical derivative in terms of the theoretical scaling of the error and, at least under the specific setups of our numerical experiments, the actual number of measurements to obtain the \pol with the same fixed accuracy.

For future work, it is interesting to investigate other derivatives such as the IR absorption intensity, which is given by a cross-derivative of the energy with respect to the electric field and atomic (normal) coordinates.
It is also fascinating to test our methods on actual quantum devices with error mitigation techniques~\cite{temme2017error, endo2018practical} as a crucial step in using near-term quantum computers in quantum chemistry calculations.

\begin{acknowledgement}
This work is supported by MEXT Quantum Leap Flagship Program (MEXT QLEAP) Grant No. JPMXS0118067394 and JPMXS0120319794.
We also acknowledge support from JST COI-NEXT program Grant No. JPMJPF2014.
WM is supported by JST PRESTO Grant No. JPMJPR191A.
\end{acknowledgement}

\section*{Supporting Information.}
The Supporting Information is provided to describe the details of the analytical formulas and numerical simulations in the main text.

%%%%% ----- %%%%%
\bibliography{bibliography.bib}

\providecommand{\latin}[1]{#1}
\makeatletter
\providecommand{\doi}
  {\begingroup\let\do\@makeother\dospecials
  \catcode`\{=1 \catcode`\}=2 \doi@aux}
\providecommand{\doi@aux}[1]{\endgroup\texttt{#1}}
\makeatother
\providecommand*\mcitethebibliography{\thebibliography}
\csname @ifundefined\endcsname{endmcitethebibliography}
  {\let\endmcitethebibliography\endthebibliography}{}
\begin{mcitethebibliography}{55}
\providecommand*\natexlab[1]{#1}
\providecommand*\mciteSetBstSublistMode[1]{}
\providecommand*\mciteSetBstMaxWidthForm[2]{}
\providecommand*\mciteBstWouldAddEndPuncttrue
  {\def\EndOfBibitem{\unskip.}}
\providecommand*\mciteBstWouldAddEndPunctfalse
  {\let\EndOfBibitem\relax}
\providecommand*\mciteSetBstMidEndSepPunct[3]{}
\providecommand*\mciteSetBstSublistLabelBeginEnd[3]{}
\providecommand*\EndOfBibitem{}
\mciteSetBstSublistMode{f}
\mciteSetBstMaxWidthForm{subitem}{(\alph{mcitesubitemcount})}
\mciteSetBstSublistLabelBeginEnd
  {\mcitemaxwidthsubitemform\space}
  {\relax}
  {\relax}

\bibitem[Nielsen and Chuang(2011)Nielsen, and Chuang]{nielsen2002quantum}
Nielsen,~M.~A.; Chuang,~I.~L. \emph{Quantum Computation and Quantum
  Information: 10th Anniversary Edition}; Cambridge University Press,
  2011\relax
\mciteBstWouldAddEndPuncttrue
\mciteSetBstMidEndSepPunct{\mcitedefaultmidpunct}
{\mcitedefaultendpunct}{\mcitedefaultseppunct}\relax
\EndOfBibitem
\bibitem[Preskill(2018)]{Preskill2018}
Preskill,~J. Quantum {C}omputing in the {NISQ} era and beyond. \emph{{Quantum}}
  \textbf{2018}, \emph{2}, 79\relax
\mciteBstWouldAddEndPuncttrue
\mciteSetBstMidEndSepPunct{\mcitedefaultmidpunct}
{\mcitedefaultendpunct}{\mcitedefaultseppunct}\relax
\EndOfBibitem
\bibitem[Arute \latin{et~al.}(2019)Arute, Arya, Babbush, Bacon, Bardin,
  Barends, Biswas, Boixo, Brandao, Buell, Burkett, Chen, Chen, Chiaro, Collins,
  Courtney, Dunsworth, Farhi, Foxen, Fowler, Gidney, Giustina, Graff, Guerin,
  Habegger, Harrigan, Hartmann, Ho, Hoffmann, Huang, Humble, Isakov, Jeffrey,
  Jiang, Kafri, Kechedzhi, Kelly, Klimov, Knysh, Korotkov, Kostritsa, Landhuis,
  Lindmark, Lucero, Lyakh, Mandr{\`a}, McClean, McEwen, Megrant, Mi,
  Michielsen, Mohseni, Mutus, Naaman, Neeley, Neill, Niu, Ostby, Petukhov,
  Platt, Quintana, Rieffel, Roushan, Rubin, Sank, Satzinger, Smelyanskiy, Sung,
  Trevithick, Vainsencher, Villalonga, White, Yao, Yeh, Zalcman, Neven, and
  Martinis]{Arute2019}
Arute,~F.; Arya,~K.; Babbush,~R.; Bacon,~D.; Bardin,~J.~C.; Barends,~R.;
  Biswas,~R.; Boixo,~S.; Brandao,~F. G. S.~L.; Buell,~D.~A.; Burkett,~B.;
  Chen,~Y.; Chen,~Z.; Chiaro,~B.; Collins,~R.; Courtney,~W.; Dunsworth,~A.;
  Farhi,~E.; Foxen,~B.; Fowler,~A.; Gidney,~C.; Giustina,~M.; Graff,~R.;
  Guerin,~K.; Habegger,~S.; Harrigan,~M.~P.; Hartmann,~M.~J.; Ho,~A.;
  Hoffmann,~M.; Huang,~T.; Humble,~T.~S.; Isakov,~S.~V.; Jeffrey,~E.;
  Jiang,~Z.; Kafri,~D.; Kechedzhi,~K.; Kelly,~J.; Klimov,~P.~V.; Knysh,~S.;
  Korotkov,~A.; Kostritsa,~F.; Landhuis,~D.; Lindmark,~M.; Lucero,~E.;
  Lyakh,~D.; Mandr{\`a},~S.; McClean,~J.~R.; McEwen,~M.; Megrant,~A.; Mi,~X.;
  Michielsen,~K.; Mohseni,~M.; Mutus,~J.; Naaman,~O.; Neeley,~M.; Neill,~C.;
  Niu,~M.~Y.; Ostby,~E.; Petukhov,~A.; Platt,~J.~C.; Quintana,~C.;
  Rieffel,~E.~G.; Roushan,~P.; Rubin,~N.~C.; Sank,~D.; Satzinger,~K.~J.;
  Smelyanskiy,~V.; Sung,~K.~J.; Trevithick,~M.~D.; Vainsencher,~A.;
  Villalonga,~B.; White,~T.; Yao,~Z.~J.; Yeh,~P.; Zalcman,~A.; Neven,~H.;
  Martinis,~J.~M. Quantum supremacy using a programmable superconducting
  processor. \emph{Nature} \textbf{2019}, \emph{574}, 505--510\relax
\mciteBstWouldAddEndPuncttrue
\mciteSetBstMidEndSepPunct{\mcitedefaultmidpunct}
{\mcitedefaultendpunct}{\mcitedefaultseppunct}\relax
\EndOfBibitem
\bibitem[Wu \latin{et~al.}(2021)Wu, Bao, Cao, Chen, Chen, Chen, Chung, Deng,
  Du, Fan, Gong, Guo, Guo, Guo, Han, Hong, Huang, Huo, Li, Li, Li, Li, Liang,
  Lin, Lin, Qian, Qiao, Rong, Su, Sun, Wang, Wang, Wu, Xu, Yan, Yang, Yang, Ye,
  Yin, Ying, Yu, Zha, Zhang, Zhang, Zhang, Zhang, Zhao, Zhao, Zhou, Zhu, Lu,
  Peng, Zhu, and Pan]{Wu2021}
Wu,~Y.; Bao,~W.-S.; Cao,~S.; Chen,~F.; Chen,~M.-C.; Chen,~X.; Chung,~T.-H.;
  Deng,~H.; Du,~Y.; Fan,~D.; Gong,~M.; Guo,~C.; Guo,~C.; Guo,~S.; Han,~L.;
  Hong,~L.; Huang,~H.-L.; Huo,~Y.-H.; Li,~L.; Li,~N.; Li,~S.; Li,~Y.;
  Liang,~F.; Lin,~C.; Lin,~J.; Qian,~H.; Qiao,~D.; Rong,~H.; Su,~H.; Sun,~L.;
  Wang,~L.; Wang,~S.; Wu,~D.; Xu,~Y.; Yan,~K.; Yang,~W.; Yang,~Y.; Ye,~Y.;
  Yin,~J.; Ying,~C.; Yu,~J.; Zha,~C.; Zhang,~C.; Zhang,~H.; Zhang,~K.;
  Zhang,~Y.; Zhao,~H.; Zhao,~Y.; Zhou,~L.; Zhu,~Q.; Lu,~C.-Y.; Peng,~C.-Z.;
  Zhu,~X.; Pan,~J.-W. Strong Quantum Computational Advantage Using a
  Superconducting Quantum Processor. \emph{Phys. Rev. Lett.} \textbf{2021},
  \emph{127}, 180501\relax
\mciteBstWouldAddEndPuncttrue
\mciteSetBstMidEndSepPunct{\mcitedefaultmidpunct}
{\mcitedefaultendpunct}{\mcitedefaultseppunct}\relax
\EndOfBibitem
\bibitem[Zhong \latin{et~al.}(2020)Zhong, Wang, Deng, Chen, Peng, Luo, Qin, Wu,
  Ding, Hu, Hu, Yang, Zhang, Li, Li, Jiang, Gan, Yang, You, Wang, Li, Liu, Lu,
  and Pan]{Zhong1460}
Zhong,~H.-S.; Wang,~H.; Deng,~Y.-H.; Chen,~M.-C.; Peng,~L.-C.; Luo,~Y.-H.;
  Qin,~J.; Wu,~D.; Ding,~X.; Hu,~Y.; Hu,~P.; Yang,~X.-Y.; Zhang,~W.-J.; Li,~H.;
  Li,~Y.; Jiang,~X.; Gan,~L.; Yang,~G.; You,~L.; Wang,~Z.; Li,~L.; Liu,~N.-L.;
  Lu,~C.-Y.; Pan,~J.-W. Quantum computational advantage using photons.
  \emph{Science} \textbf{2020}, \emph{370}, 1460--1463\relax
\mciteBstWouldAddEndPuncttrue
\mciteSetBstMidEndSepPunct{\mcitedefaultmidpunct}
{\mcitedefaultendpunct}{\mcitedefaultseppunct}\relax
\EndOfBibitem
\bibitem[Madsen \latin{et~al.}(2022)Madsen, Laudenbach, Askarani, Rortais,
  Vincent, Bulmer, Miatto, Neuhaus, Helt, Collins, Lita, Gerrits, Nam, Vaidya,
  Menotti, Dhand, Vernon, Quesada, and Lavoie]{Madsen2022}
Madsen,~L.~S.; Laudenbach,~F.; Askarani,~M.~F.; Rortais,~F.; Vincent,~T.;
  Bulmer,~J. F.~F.; Miatto,~F.~M.; Neuhaus,~L.; Helt,~L.~G.; Collins,~M.~J.;
  Lita,~A.~E.; Gerrits,~T.; Nam,~S.~W.; Vaidya,~V.~D.; Menotti,~M.; Dhand,~I.;
  Vernon,~Z.; Quesada,~N.; Lavoie,~J. Quantum computational advantage with a
  programmable photonic processor. \emph{Nature} \textbf{2022}, \emph{606},
  75--81\relax
\mciteBstWouldAddEndPuncttrue
\mciteSetBstMidEndSepPunct{\mcitedefaultmidpunct}
{\mcitedefaultendpunct}{\mcitedefaultseppunct}\relax
\EndOfBibitem
\bibitem[Shor(1997)]{shor1999polynomial}
Shor,~P.~W. Polynomial-Time Algorithms for Prime Factorization and Discrete
  Logarithms on a Quantum Computer. \emph{SIAM J. Comput.} \textbf{1997},
  \emph{26}, 1484^^e2^^80^^931509\relax
\mciteBstWouldAddEndPuncttrue
\mciteSetBstMidEndSepPunct{\mcitedefaultmidpunct}
{\mcitedefaultendpunct}{\mcitedefaultseppunct}\relax
\EndOfBibitem
\bibitem[Grover(1996)]{grover1996fast}
Grover,~L.~K. A Fast Quantum Mechanical Algorithm for Database Search.
  Proceedings of the Twenty-Eighth Annual ACM Symposium on Theory of Computing.
  New York, NY, USA, 1996; p 212^^e2^^80^^93219\relax
\mciteBstWouldAddEndPuncttrue
\mciteSetBstMidEndSepPunct{\mcitedefaultmidpunct}
{\mcitedefaultendpunct}{\mcitedefaultseppunct}\relax
\EndOfBibitem
\bibitem[Kitaev(1995)]{kitaev1995quantum}
Kitaev,~A.~Y. Quantum measurements and the Abelian stabilizer problem.
  \emph{arXiv preprint quant-ph/9511026} \textbf{1995}, \relax
\mciteBstWouldAddEndPunctfalse
\mciteSetBstMidEndSepPunct{\mcitedefaultmidpunct}
{}{\mcitedefaultseppunct}\relax
\EndOfBibitem
\bibitem[Cleve \latin{et~al.}(1998)Cleve, Ekert, Macchiavello, and
  Mosca]{Cleve1998}
Cleve,~R.; Ekert,~A.; Macchiavello,~C.; Mosca,~M. Quantum algorithms revisited.
  \emph{Proc. R. Soc. London, Ser. A} \textbf{1998}, \emph{454}, 339--354\relax
\mciteBstWouldAddEndPuncttrue
\mciteSetBstMidEndSepPunct{\mcitedefaultmidpunct}
{\mcitedefaultendpunct}{\mcitedefaultseppunct}\relax
\EndOfBibitem
\bibitem[Peruzzo \latin{et~al.}(2014)Peruzzo, McClean, Shadbolt, Yung, Zhou,
  Love, Aspuru-Guzik, and O’Brien]{peruzzo2014variational}
Peruzzo,~A.; McClean,~J.; Shadbolt,~P.; Yung,~M.-H.; Zhou,~X.-Q.; Love,~P.~J.;
  Aspuru-Guzik,~A.; O’Brien,~J.~L. A variational eigenvalue solver on a
  photonic quantum processor. \emph{Nat. Commun.} \textbf{2014}, \emph{5}\relax
\mciteBstWouldAddEndPuncttrue
\mciteSetBstMidEndSepPunct{\mcitedefaultmidpunct}
{\mcitedefaultendpunct}{\mcitedefaultseppunct}\relax
\EndOfBibitem
\bibitem[McClean \latin{et~al.}(2016)McClean, Romero, Babbush, and
  Aspuru-Guzik]{mcclean2016theory}
McClean,~J.~R.; Romero,~J.; Babbush,~R.; Aspuru-Guzik,~A. The theory of
  variational hybrid quantum-classical algorithms. \emph{New J. Phys.}
  \textbf{2016}, \emph{18}, 023023\relax
\mciteBstWouldAddEndPuncttrue
\mciteSetBstMidEndSepPunct{\mcitedefaultmidpunct}
{\mcitedefaultendpunct}{\mcitedefaultseppunct}\relax
\EndOfBibitem
\bibitem[Tilly \latin{et~al.}(2021)Tilly, Chen, Cao, Picozzi, Setia, Li, Grant,
  Wossnig, Rungger, Booth, \latin{et~al.} others]{tilly2021variational}
Tilly,~J.; Chen,~H.; Cao,~S.; Picozzi,~D.; Setia,~K.; Li,~Y.; Grant,~E.;
  Wossnig,~L.; Rungger,~I.; Booth,~G.~H., \latin{et~al.}  The variational
  quantum eigensolver: a review of methods and best practices. \emph{arXiv
  preprint arXiv:2111.05176} \textbf{2021}, \relax
\mciteBstWouldAddEndPunctfalse
\mciteSetBstMidEndSepPunct{\mcitedefaultmidpunct}
{}{\mcitedefaultseppunct}\relax
\EndOfBibitem
\bibitem[Kandala \latin{et~al.}(2017)Kandala, Mezzacapo, Temme, Takita, Brink,
  Chow, and Gambetta]{kandala2017hardware}
Kandala,~A.; Mezzacapo,~A.; Temme,~K.; Takita,~M.; Brink,~M.; Chow,~J.~M.;
  Gambetta,~J.~M. Hardware-efficient variational quantum eigensolver for small
  molecules and quantum magnets. \emph{Nature} \textbf{2017}, \emph{549},
  242^^e2^^80^^93246\relax
\mciteBstWouldAddEndPuncttrue
\mciteSetBstMidEndSepPunct{\mcitedefaultmidpunct}
{\mcitedefaultendpunct}{\mcitedefaultseppunct}\relax
\EndOfBibitem
\bibitem[Colless \latin{et~al.}(2018)Colless, Ramasesh, Dahlen, Blok,
  Kimchi-Schwartz, McClean, Carter, de~Jong, and
  Siddiqi]{colless2018computation}
Colless,~J.~I.; Ramasesh,~V.~V.; Dahlen,~D.; Blok,~M.~S.;
  Kimchi-Schwartz,~M.~E.; McClean,~J.~R.; Carter,~J.; de~Jong,~W.~A.;
  Siddiqi,~I. Computation of Molecular Spectra on a Quantum Processor with an
  Error-Resilient Algorithm. \emph{Phys. Rev. X} \textbf{2018}, \emph{8},
  011021\relax
\mciteBstWouldAddEndPuncttrue
\mciteSetBstMidEndSepPunct{\mcitedefaultmidpunct}
{\mcitedefaultendpunct}{\mcitedefaultseppunct}\relax
\EndOfBibitem
\bibitem[Kandala \latin{et~al.}(2019)Kandala, Temme, C{\'o}rcoles, Mezzacapo,
  Chow, and Gambetta]{kandala2019error}
Kandala,~A.; Temme,~K.; C{\'o}rcoles,~A.~D.; Mezzacapo,~A.; Chow,~J.~M.;
  Gambetta,~J.~M. Error mitigation extends the computational reach of a noisy
  quantum processor. \emph{Nature} \textbf{2019}, \emph{567}, 491--495\relax
\mciteBstWouldAddEndPuncttrue
\mciteSetBstMidEndSepPunct{\mcitedefaultmidpunct}
{\mcitedefaultendpunct}{\mcitedefaultseppunct}\relax
\EndOfBibitem
\bibitem[McArdle \latin{et~al.}(2020)McArdle, Endo, Aspuru-Guzik, Benjamin, and
  Yuan]{mcardle2018quantum}
McArdle,~S.; Endo,~S.; Aspuru-Guzik,~A.; Benjamin,~S.~C.; Yuan,~X. Quantum
  computational chemistry. \emph{Rev. Mod. Phys.} \textbf{2020}, \emph{92},
  015003\relax
\mciteBstWouldAddEndPuncttrue
\mciteSetBstMidEndSepPunct{\mcitedefaultmidpunct}
{\mcitedefaultendpunct}{\mcitedefaultseppunct}\relax
\EndOfBibitem
\bibitem[Cao \latin{et~al.}(2019)Cao, Romero, Olson, Degroote, Johnson,
  Kieferov{\'a}, Kivlichan, Menke, Peropadre, Sawaya, Sim, Veis, and
  Aspuru-Guzik]{Cao2018}
Cao,~Y.; Romero,~J.; Olson,~J.~P.; Degroote,~M.; Johnson,~P.~D.;
  Kieferov{\'a},~M.; Kivlichan,~I.~D.; Menke,~T.; Peropadre,~B.; Sawaya,~N.
  P.~D.; Sim,~S.; Veis,~L.; Aspuru-Guzik,~A. Quantum Chemistry in the Age of
  Quantum Computing. \emph{Chem. Rev.} \textbf{2019}, \emph{119},
  10856--10915\relax
\mciteBstWouldAddEndPuncttrue
\mciteSetBstMidEndSepPunct{\mcitedefaultmidpunct}
{\mcitedefaultendpunct}{\mcitedefaultseppunct}\relax
\EndOfBibitem
\bibitem[Mitarai \latin{et~al.}(2020)Mitarai, Nakagawa, and
  Mizukami]{mitarai2020theory}
Mitarai,~K.; Nakagawa,~Y.~O.; Mizukami,~W. Theory of analytical energy
  derivatives for the variational quantum eigensolver. \emph{Physical Review
  Research} \textbf{2020}, \emph{2}, 013129\relax
\mciteBstWouldAddEndPuncttrue
\mciteSetBstMidEndSepPunct{\mcitedefaultmidpunct}
{\mcitedefaultendpunct}{\mcitedefaultseppunct}\relax
\EndOfBibitem
\bibitem[Parrish \latin{et~al.}(2019)Parrish, Hohenstein, McMahon, and
  Martinez]{parrish2019hybrid}
Parrish,~R.~M.; Hohenstein,~E.~G.; McMahon,~P.~L.; Martinez,~T.~J. Hybrid
  quantum/classical derivative theory: Analytical gradients and excited-state
  dynamics for the multistate contracted variational quantum eigensolver.
  \emph{arXiv preprint arXiv:1906.08728} \textbf{2019}, \relax
\mciteBstWouldAddEndPunctfalse
\mciteSetBstMidEndSepPunct{\mcitedefaultmidpunct}
{}{\mcitedefaultseppunct}\relax
\EndOfBibitem
\bibitem[Tamiya \latin{et~al.}(2021)Tamiya, Koh, and Nakagawa]{tamiya2021}
Tamiya,~S.; Koh,~S.; Nakagawa,~Y.~O. Calculating nonadiabatic couplings and
  Berry's phase by variational quantum eigensolvers. \emph{Phys. Rev. Research}
  \textbf{2021}, \emph{3}, 023244\relax
\mciteBstWouldAddEndPuncttrue
\mciteSetBstMidEndSepPunct{\mcitedefaultmidpunct}
{\mcitedefaultendpunct}{\mcitedefaultseppunct}\relax
\EndOfBibitem
\bibitem[Parrish \latin{et~al.}(2021)Parrish, Anselmetti, and
  Gogolin]{parrish2021analytical}
Parrish,~R.~M.; Anselmetti,~G.-L.~R.; Gogolin,~C. Analytical Ground-and
  Excited-State Gradients for Molecular Electronic Structure Theory from Hybrid
  Quantum/Classical Methods. \emph{arXiv preprint arXiv:2110.05040}
  \textbf{2021}, \relax
\mciteBstWouldAddEndPunctfalse
\mciteSetBstMidEndSepPunct{\mcitedefaultmidpunct}
{}{\mcitedefaultseppunct}\relax
\EndOfBibitem
\bibitem[Yalouz \latin{et~al.}(2022)Yalouz, Koridon, Senjean, Lasorne, Buda,
  and Visscher]{yalouz2022analytical}
Yalouz,~S.; Koridon,~E.; Senjean,~B.; Lasorne,~B.; Buda,~F.; Visscher,~L.
  Analytical Nonadiabatic Couplings and Gradients within the State-Averaged
  Orbital-Optimized Variational Quantum Eigensolver. \emph{Journal of Chemical
  Theory and Computation} \textbf{2022}, \emph{18}, 776--794, PMID:
  35029988\relax
\mciteBstWouldAddEndPuncttrue
\mciteSetBstMidEndSepPunct{\mcitedefaultmidpunct}
{\mcitedefaultendpunct}{\mcitedefaultseppunct}\relax
\EndOfBibitem
\bibitem[Omiya \latin{et~al.}(2022)Omiya, Nakagawa, Koh, Mizukami, Gao, and
  Kobayashi]{omiya2022}
Omiya,~K.; Nakagawa,~Y.~O.; Koh,~S.; Mizukami,~W.; Gao,~Q.; Kobayashi,~T.
  Analytical Energy Gradient for State-Averaged Orbital-Optimized Variational
  Quantum Eigensolvers and Its Application to a Photochemical Reaction.
  \emph{Journal of Chemical Theory and Computation} \textbf{2022}, \emph{18},
  741--748\relax
\mciteBstWouldAddEndPuncttrue
\mciteSetBstMidEndSepPunct{\mcitedefaultmidpunct}
{\mcitedefaultendpunct}{\mcitedefaultseppunct}\relax
\EndOfBibitem
\bibitem[Hohenstein \latin{et~al.}(2022)Hohenstein, Oumarou, Al-Saadon,
  Anselmetti, Scheurer, Gogolin, and Parrish]{hohenstein2022efficient}
Hohenstein,~E.~G.; Oumarou,~O.; Al-Saadon,~R.; Anselmetti,~G.-L.~R.;
  Scheurer,~M.; Gogolin,~C.; Parrish,~R.~M. Efficient Quantum Analytic Nuclear
  Gradients with Double Factorization. \emph{arXiv preprint arXiv:2207.13144}
  \textbf{2022}, \relax
\mciteBstWouldAddEndPunctfalse
\mciteSetBstMidEndSepPunct{\mcitedefaultmidpunct}
{}{\mcitedefaultseppunct}\relax
\EndOfBibitem
\bibitem[O’Brien \latin{et~al.}(2019)O’Brien, Senjean, Sagastizabal,
  Bonet-Monroig, Dutkiewicz, Buda, DiCarlo, and Visscher]{o2019calculating}
O’Brien,~T.~E.; Senjean,~B.; Sagastizabal,~R.; Bonet-Monroig,~X.;
  Dutkiewicz,~A.; Buda,~F.; DiCarlo,~L.; Visscher,~L. Calculating energy
  derivatives for quantum chemistry on a quantum computer. \emph{npj Quantum
  Information} \textbf{2019}, \emph{5}, 1--12\relax
\mciteBstWouldAddEndPuncttrue
\mciteSetBstMidEndSepPunct{\mcitedefaultmidpunct}
{\mcitedefaultendpunct}{\mcitedefaultseppunct}\relax
\EndOfBibitem
\bibitem[O'Brien \latin{et~al.}(2021)O'Brien, Streif, Rubin, Santagati, Su,
  Huggins, Goings, Moll, Kyoseva, Degroote, \latin{et~al.}
  others]{o2021efficient}
O'Brien,~T.~E.; Streif,~M.; Rubin,~N.~C.; Santagati,~R.; Su,~Y.;
  Huggins,~W.~J.; Goings,~J.~J.; Moll,~N.; Kyoseva,~E.; Degroote,~M.,
  \latin{et~al.}  Efficient quantum computation of molecular forces and other
  energy gradients. \emph{arXiv preprint arXiv:2111.12437} \textbf{2021},
  \relax
\mciteBstWouldAddEndPunctfalse
\mciteSetBstMidEndSepPunct{\mcitedefaultmidpunct}
{}{\mcitedefaultseppunct}\relax
\EndOfBibitem
\bibitem[Takeshita \latin{et~al.}(2020)Takeshita, Rubin, Jiang, Lee, Babbush,
  and McClean]{Takeshita2020}
Takeshita,~T.; Rubin,~N.~C.; Jiang,~Z.; Lee,~E.; Babbush,~R.; McClean,~J.~R.
  Increasing the Representation Accuracy of Quantum Simulations of Chemistry
  without Extra Quantum Resources. \emph{Phys. Rev. X} \textbf{2020},
  \emph{10}, 011004\relax
\mciteBstWouldAddEndPuncttrue
\mciteSetBstMidEndSepPunct{\mcitedefaultmidpunct}
{\mcitedefaultendpunct}{\mcitedefaultseppunct}\relax
\EndOfBibitem
\bibitem[Mizukami \latin{et~al.}(2020)Mizukami, Mitarai, Nakagawa, Yamamoto,
  Yan, and Ohnishi]{Mizukami2020}
Mizukami,~W.; Mitarai,~K.; Nakagawa,~Y.~O.; Yamamoto,~T.; Yan,~T.;
  Ohnishi,~Y.-y. Orbital optimized unitary coupled cluster theory for quantum
  computer. \emph{Phys. Rev. Res.} \textbf{2020}, \emph{2}, 033421\relax
\mciteBstWouldAddEndPuncttrue
\mciteSetBstMidEndSepPunct{\mcitedefaultmidpunct}
{\mcitedefaultendpunct}{\mcitedefaultseppunct}\relax
\EndOfBibitem
\bibitem[Sokolov \latin{et~al.}(2020)Sokolov, Barkoutsos, Ollitrault,
  Greenberg, Rice, Pistoia, and Tavernelli]{Sokolov2020JCP}
Sokolov,~I.~O.; Barkoutsos,~P.~K.; Ollitrault,~P.~J.; Greenberg,~D.; Rice,~J.;
  Pistoia,~M.; Tavernelli,~I. Quantum orbital-optimized unitary coupled cluster
  methods in the strongly correlated regime: Can quantum algorithms outperform
  their classical equivalents? \emph{J. Chem. Phys.} \textbf{2020}, \emph{152},
  124107\relax
\mciteBstWouldAddEndPuncttrue
\mciteSetBstMidEndSepPunct{\mcitedefaultmidpunct}
{\mcitedefaultendpunct}{\mcitedefaultseppunct}\relax
\EndOfBibitem
\bibitem[Szalay \latin{et~al.}(2012)Szalay, M{\" u}ller, Gidofalvi, Lischka,
  and Shepard]{Szalay2012}
Szalay,~P.~G.; M{\" u}ller,~T.; Gidofalvi,~G.; Lischka,~H.; Shepard,~R.
  Multiconfiguration Self-Consistent Field and Multireference Configuration
  Interaction Methods and Applications. \emph{Chem. Rev.} \textbf{2012},
  \emph{112}, 108--181, PMID: 22204633\relax
\mciteBstWouldAddEndPuncttrue
\mciteSetBstMidEndSepPunct{\mcitedefaultmidpunct}
{\mcitedefaultendpunct}{\mcitedefaultseppunct}\relax
\EndOfBibitem
\bibitem[Roos \latin{et~al.}(2016)Roos, Lindh, Malmqvist, Veryazov, and
  Widmark]{Roos2016multiconfigurational}
Roos,~B.; Lindh,~R.; Malmqvist,~P.; Veryazov,~V.; Widmark,~P.
  \emph{Multiconfigurational Quantum Chemistry}; John Wiley \& Sons, Ltd,
  2016\relax
\mciteBstWouldAddEndPuncttrue
\mciteSetBstMidEndSepPunct{\mcitedefaultmidpunct}
{\mcitedefaultendpunct}{\mcitedefaultseppunct}\relax
\EndOfBibitem
\bibitem[Bozkaya \latin{et~al.}(2011)Bozkaya, Turney, Yamaguchi, Schaefer, and
  Sherrill]{Bozkaya2011}
Bozkaya,~U.; Turney,~J.~M.; Yamaguchi,~Y.; Schaefer,~H.~F.; Sherrill,~C.~D.
  Quadratically convergent algorithm for orbital optimization in the
  orbital-optimized coupled-cluster doubles method and in orbital-optimized
  second-order M^^c3^^b8ller-Plesset perturbation theory. \emph{The Journal of
  Chemical Physics} \textbf{2011}, \emph{135}, 104103\relax
\mciteBstWouldAddEndPuncttrue
\mciteSetBstMidEndSepPunct{\mcitedefaultmidpunct}
{\mcitedefaultendpunct}{\mcitedefaultseppunct}\relax
\EndOfBibitem
\bibitem[Mitarai \latin{et~al.}(2018)Mitarai, Negoro, Kitagawa, and
  Fujii]{Mitarai2018}
Mitarai,~K.; Negoro,~M.; Kitagawa,~M.; Fujii,~K. Quantum circuit learning.
  \emph{Phys. Rev. A} \textbf{2018}, \emph{98}, 032309\relax
\mciteBstWouldAddEndPuncttrue
\mciteSetBstMidEndSepPunct{\mcitedefaultmidpunct}
{\mcitedefaultendpunct}{\mcitedefaultseppunct}\relax
\EndOfBibitem
\bibitem[Schuld \latin{et~al.}(2019)Schuld, Bergholm, Gogolin, Izaac, and
  Killoran]{Schuld2019}
Schuld,~M.; Bergholm,~V.; Gogolin,~C.; Izaac,~J.; Killoran,~N. Evaluating
  analytic gradients on quantum hardware. \emph{Phys. Rev. A} \textbf{2019},
  \emph{99}, 032331\relax
\mciteBstWouldAddEndPuncttrue
\mciteSetBstMidEndSepPunct{\mcitedefaultmidpunct}
{\mcitedefaultendpunct}{\mcitedefaultseppunct}\relax
\EndOfBibitem
\bibitem[Izmaylov \latin{et~al.}(2021)Izmaylov, Lang, and
  Yen]{izmaylov2021analytic}
Izmaylov,~A.~F.; Lang,~R.~A.; Yen,~T.-C. Analytic gradients in variational
  quantum algorithms: Algebraic extensions of the parameter-shift rule to
  general unitary transformations. \emph{arXiv preprint arXiv:2107.08131}
  \textbf{2021}, \relax
\mciteBstWouldAddEndPunctfalse
\mciteSetBstMidEndSepPunct{\mcitedefaultmidpunct}
{}{\mcitedefaultseppunct}\relax
\EndOfBibitem
\bibitem[Helgaker and J{\o}rgensen(1989)Helgaker, and
  J{\o}rgensen]{Helgaker1989}
Helgaker,~T.; J{\o}rgensen,~P. Configuration-interaction energy derivatives in
  a fully variational formulation. \emph{Theor. Chim. Acta} \textbf{1989},
  \emph{75}, 111--127\relax
\mciteBstWouldAddEndPuncttrue
\mciteSetBstMidEndSepPunct{\mcitedefaultmidpunct}
{\mcitedefaultendpunct}{\mcitedefaultseppunct}\relax
\EndOfBibitem
\bibitem[{Editor: Russell D. Johnson III}(2022)]{cccbdb}
{Editor: Russell D. Johnson III}, {NIST Computational Chemistry Comparison and
  Benchmark Database}, {NIST Standard Reference Database Number 101}, {Release
  22, May 2022}. 2022; \url{http://cccbdb.nist.gov/}\relax
\mciteBstWouldAddEndPuncttrue
\mciteSetBstMidEndSepPunct{\mcitedefaultmidpunct}
{\mcitedefaultendpunct}{\mcitedefaultseppunct}\relax
\EndOfBibitem
\bibitem[Sun \latin{et~al.}(2018)Sun, Berkelbach, Blunt, Booth, Guo, Li, Liu,
  McClain, Sayfutyarova, Sharma, Wouters, and Chan]{Sun2018_pyscf}
Sun,~Q.; Berkelbach,~T.~C.; Blunt,~N.~S.; Booth,~G.~H.; Guo,~S.; Li,~Z.;
  Liu,~J.; McClain,~J.~D.; Sayfutyarova,~E.~R.; Sharma,~S.; Wouters,~S.;
  Chan,~G. K.-L. {PySCF}: the {P}ython-based simulations of chemistry
  framework. \emph{WIREs Comput. Mol. Sci.} \textbf{2018}, \emph{8},
  e1340\relax
\mciteBstWouldAddEndPuncttrue
\mciteSetBstMidEndSepPunct{\mcitedefaultmidpunct}
{\mcitedefaultendpunct}{\mcitedefaultseppunct}\relax
\EndOfBibitem
\bibitem[Anand \latin{et~al.}(2022)Anand, Schleich, Alperin-Lea, Jensen, Sim,
  D^^c3^^adaz-Tinoco, Kottmann, Degroote, Izmaylov, and
  Aspuru-Guzik]{anand2022}
Anand,~A.; Schleich,~P.; Alperin-Lea,~S.; Jensen,~P. W.~K.; Sim,~S.;
  D^^c3^^adaz-Tinoco,~M.; Kottmann,~J.~S.; Degroote,~M.; Izmaylov,~A.~F.;
  Aspuru-Guzik,~A. A quantum computing view on unitary coupled cluster theory.
  \emph{Chem. Soc. Rev.} \textbf{2022}, \emph{51}, 1659--1684\relax
\mciteBstWouldAddEndPuncttrue
\mciteSetBstMidEndSepPunct{\mcitedefaultmidpunct}
{\mcitedefaultendpunct}{\mcitedefaultseppunct}\relax
\EndOfBibitem
\bibitem[Jordan and Wigner(1928)Jordan, and Wigner]{Jordan1928}
Jordan,~P.; Wigner,~E. {\"U}ber das Paulische {\"A}quivalenzverbot.
  \emph{Zeitschrift f{\"u}r Physik} \textbf{1928}, \emph{47}, 631--651\relax
\mciteBstWouldAddEndPuncttrue
\mciteSetBstMidEndSepPunct{\mcitedefaultmidpunct}
{\mcitedefaultendpunct}{\mcitedefaultseppunct}\relax
\EndOfBibitem
\bibitem[Virtanen \latin{et~al.}(2020)Virtanen, Gommers, Oliphant, Haberland,
  Reddy, Cournapeau, Burovski, Peterson, Weckesser, Bright, {van der Walt},
  Brett, Wilson, Millman, Mayorov, Nelson, Jones, Kern, Larson, Carey, Polat,
  Feng, Moore, {VanderPlas}, Laxalde, Perktold, Cimrman, Henriksen, Quintero,
  Harris, Archibald, Ribeiro, Pedregosa, {van Mulbregt}, and {SciPy 1.0
  Contributors}]{SciPy2020}
Virtanen,~P.; Gommers,~R.; Oliphant,~T.~E.; Haberland,~M.; Reddy,~T.;
  Cournapeau,~D.; Burovski,~E.; Peterson,~P.; Weckesser,~W.; Bright,~J.; {van
  der Walt},~S.~J.; Brett,~M.; Wilson,~J.; Millman,~K.~J.; Mayorov,~N.;
  Nelson,~A. R.~J.; Jones,~E.; Kern,~R.; Larson,~E.; Carey,~C.~J.;
  Polat,~{\.I}.; Feng,~Y.; Moore,~E.~W.; {VanderPlas},~J.; Laxalde,~D.;
  Perktold,~J.; Cimrman,~R.; Henriksen,~I.; Quintero,~E.~A.; Harris,~C.~R.;
  Archibald,~A.~M.; Ribeiro,~A.~H.; Pedregosa,~F.; {van Mulbregt},~P.; {SciPy
  1.0 Contributors}, {{SciPy} 1.0: Fundamental Algorithms for Scientific
  Computing in Python}. \emph{Nature Methods} \textbf{2020}, \emph{17},
  261--272\relax
\mciteBstWouldAddEndPuncttrue
\mciteSetBstMidEndSepPunct{\mcitedefaultmidpunct}
{\mcitedefaultendpunct}{\mcitedefaultseppunct}\relax
\EndOfBibitem
\bibitem[Suzuki \latin{et~al.}(2021)Suzuki, Kawase, Masumura, Hiraga, Nakadai,
  Chen, Nakanishi, Mitarai, Imai, Tamiya, Yamamoto, Yan, Kawakubo, Nakagawa,
  Ibe, Zhang, Yamashita, Yoshimura, Hayashi, and Fujii]{Suzuki2021qulacsfast}
Suzuki,~Y.; Kawase,~Y.; Masumura,~Y.; Hiraga,~Y.; Nakadai,~M.; Chen,~J.;
  Nakanishi,~K.~M.; Mitarai,~K.; Imai,~R.; Tamiya,~S.; Yamamoto,~T.; Yan,~T.;
  Kawakubo,~T.; Nakagawa,~Y.~O.; Ibe,~Y.; Zhang,~Y.; Yamashita,~H.;
  Yoshimura,~H.; Hayashi,~A.; Fujii,~K. Qulacs: a fast and versatile quantum
  circuit simulator for research purpose. \emph{{Quantum}} \textbf{2021},
  \emph{5}, 559\relax
\mciteBstWouldAddEndPuncttrue
\mciteSetBstMidEndSepPunct{\mcitedefaultmidpunct}
{\mcitedefaultendpunct}{\mcitedefaultseppunct}\relax
\EndOfBibitem
\bibitem[Aidas \latin{et~al.}(2014)Aidas, Angeli, Bak, Bakken, Bast, Boman,
  Christiansen, Cimiraglia, Coriani, Dahle, Dalskov, Ekstr\"{o}m, Enevoldsen,
  Eriksen, Ettenhuber, Fern\'{a}ndez, Ferrighi, Fliegl, Frediani, Hald,
  Halkier, H\"{a}ttig, Heiberg, Helgaker, Hennum, Hettema, Hjerten\ae{}s,
  H\o{}st, H\o{}yvik, Iozzi, Jans\'{i}k, Jensen, Jonsson, J\o{}rgensen,
  Kauczor, Kirpekar, Kj\ae{}rgaard, Klopper, Knecht, Kobayashi, Koch, Kongsted,
  Krapp, Kristensen, Ligabue, Lutn\ae{}s, Melo, Mikkelsen, Myhre, Neiss,
  Nielsen, Norman, Olsen, Olsen, Osted, Packer, Pawlowski, Pedersen, Provasi,
  Reine, Rinkevicius, Ruden, Ruud, Rybkin, Sa\l{}ek, Samson, de~Mer\'{a}s,
  Saue, Sauer, Schimmelpfennig, Sneskov, Steindal, Sylvester-Hvid, Taylor,
  Teale, Tellgren, Tew, Thorvaldsen, Th\o{}gersen, Vahtras, Watson, Wilson,
  Ziolkowski, and \AA{}gren]{daltonpaper}
Aidas,~K.; Angeli,~C.; Bak,~K.~L.; Bakken,~V.; Bast,~R.; Boman,~L.;
  Christiansen,~O.; Cimiraglia,~R.; Coriani,~S.; Dahle,~P.; Dalskov,~E.~K.;
  Ekstr\"{o}m,~U.; Enevoldsen,~T.; Eriksen,~J.~J.; Ettenhuber,~P.;
  Fern\'{a}ndez,~B.; Ferrighi,~L.; Fliegl,~H.; Frediani,~L.; Hald,~K.;
  Halkier,~A.; H\"{a}ttig,~C.; Heiberg,~H.; Helgaker,~T.; Hennum,~A.~C.;
  Hettema,~H.; Hjerten\ae{}s,~E.; H\o{}st,~S.; H\o{}yvik,~I.-M.; Iozzi,~M.~F.;
  Jans\'{i}k,~B.; Jensen,~H. J.~{\relax Aa}.; Jonsson,~D.; J\o{}rgensen,~P.;
  Kauczor,~J.; Kirpekar,~S.; Kj\ae{}rgaard,~T.; Klopper,~W.; Knecht,~S.;
  Kobayashi,~R.; Koch,~H.; Kongsted,~J.; Krapp,~A.; Kristensen,~K.;
  Ligabue,~A.; Lutn\ae{}s,~O.~B.; Melo,~J.~I.; Mikkelsen,~K.~V.; Myhre,~R.~H.;
  Neiss,~C.; Nielsen,~C.~B.; Norman,~P.; Olsen,~J.; Olsen,~J. M.~H.; Osted,~A.;
  Packer,~M.~J.; Pawlowski,~F.; Pedersen,~T.~B.; Provasi,~P.~F.; Reine,~S.;
  Rinkevicius,~Z.; Ruden,~T.~A.; Ruud,~K.; Rybkin,~V.~V.; Sa\l{}ek,~P.;
  Samson,~C. C.~M.; de~Mer\'{a}s,~A.~S.; Saue,~T.; Sauer,~S. P.~A.;
  Schimmelpfennig,~B.; Sneskov,~K.; Steindal,~A.~H.; Sylvester-Hvid,~K.~O.;
  Taylor,~P.~R.; Teale,~A.~M.; Tellgren,~E.~I.; Tew,~D.~P.; Thorvaldsen,~A.~J.;
  Th\o{}gersen,~L.; Vahtras,~O.; Watson,~M.~A.; Wilson,~D. J.~D.;
  Ziolkowski,~M.; \AA{}gren,~H. {The Dalton quantum chemistry program system}.
  \emph{WIREs Comput.~Mol.~Sci.} \textbf{2014}, \emph{4}, 269--284\relax
\mciteBstWouldAddEndPuncttrue
\mciteSetBstMidEndSepPunct{\mcitedefaultmidpunct}
{\mcitedefaultendpunct}{\mcitedefaultseppunct}\relax
\EndOfBibitem
\bibitem[Telesca \latin{et~al.}(2001)Telesca, Bolink, Yunoki, Hadziioannou,
  Van~Duijnen, Snijders, Jonkman, and Sawatzky]{Telesca2001}
Telesca,~R.; Bolink,~H.; Yunoki,~S.; Hadziioannou,~G.; Van~Duijnen,~P.~T.;
  Snijders,~J.~G.; Jonkman,~H.~T.; Sawatzky,~G.~A. Density-functional study of
  the evolution of the electronic structure of oligomers of thiophene: Towards
  a model Hamiltonian. \emph{Phys. Rev. B} \textbf{2001}, \emph{63},
  155112\relax
\mciteBstWouldAddEndPuncttrue
\mciteSetBstMidEndSepPunct{\mcitedefaultmidpunct}
{\mcitedefaultendpunct}{\mcitedefaultseppunct}\relax
\EndOfBibitem
\bibitem[{David R. Lide (ed.)}(1995)]{crc76th}
{David R. Lide (ed.)}, \emph{CRC Handbook of Chemistry and Physics, 76th
  Edition}; Taylor \& Francis Inc., 1995\relax
\mciteBstWouldAddEndPuncttrue
\mciteSetBstMidEndSepPunct{\mcitedefaultmidpunct}
{\mcitedefaultendpunct}{\mcitedefaultseppunct}\relax
\EndOfBibitem
\bibitem[{David R. Lide (ed.)}(2007)]{crc88th}
{David R. Lide (ed.)}, \emph{CRC Handbook of Chemistry and Physics, 88th
  Edition}; Taylor \& Francis Inc., 2007\relax
\mciteBstWouldAddEndPuncttrue
\mciteSetBstMidEndSepPunct{\mcitedefaultmidpunct}
{\mcitedefaultendpunct}{\mcitedefaultseppunct}\relax
\EndOfBibitem
\bibitem[{S. Budavari (ed.)}(1996)]{merck1996}
{S. Budavari (ed.)}, \emph{Merck Index : An Encyclopedia of Drugs, Chemicals
  and Biologicals}; CRC Press, 1996\relax
\mciteBstWouldAddEndPuncttrue
\mciteSetBstMidEndSepPunct{\mcitedefaultmidpunct}
{\mcitedefaultendpunct}{\mcitedefaultseppunct}\relax
\EndOfBibitem
\bibitem[{Ernest W. Flick^^c2^^a0(ed.)}(1985)]{industrial1985}
{Ernest W. Flick^^c2^^a0(ed.)}, \emph{Industrial Solvents Handbook, 3rd
  Edition}; Noyes Publications, 1985\relax
\mciteBstWouldAddEndPuncttrue
\mciteSetBstMidEndSepPunct{\mcitedefaultmidpunct}
{\mcitedefaultendpunct}{\mcitedefaultseppunct}\relax
\EndOfBibitem
\bibitem[Horn and Johnson(2012)Horn, and Johnson]{horn2012matrix}
Horn,~R.; Johnson,~C. \emph{Matrix Analysis}; Cambridge University Press,
  2012\relax
\mciteBstWouldAddEndPuncttrue
\mciteSetBstMidEndSepPunct{\mcitedefaultmidpunct}
{\mcitedefaultendpunct}{\mcitedefaultseppunct}\relax
\EndOfBibitem
\bibitem[Kohda \latin{et~al.}(2022)Kohda, Imai, Kanno, Mitarai, Mizukami, and
  Nakagawa]{Kohda2022}
Kohda,~M.; Imai,~R.; Kanno,~K.; Mitarai,~K.; Mizukami,~W.; Nakagawa,~Y.~O.
  Quantum expectation-value estimation by computational basis sampling.
  \emph{Phys. Rev. Res.} \textbf{2022}, \emph{4}, 033173\relax
\mciteBstWouldAddEndPuncttrue
\mciteSetBstMidEndSepPunct{\mcitedefaultmidpunct}
{\mcitedefaultendpunct}{\mcitedefaultseppunct}\relax
\EndOfBibitem
\bibitem[Huang \latin{et~al.}(2022)Huang, Cai, Li, Ge, Hou, Li, Liu, Shi, Chen,
  Zheng, Xu, Liu, Li, Fan, and Fang]{huang2022simulating}
Huang,~K.; Cai,~X.; Li,~H.; Ge,~Z.-Y.; Hou,~R.; Li,~H.; Liu,~T.; Shi,~Y.;
  Chen,~C.; Zheng,~D.; Xu,~K.; Liu,~Z.-B.; Li,~Z.; Fan,~H.; Fang,~W.-H.
  Variational Quantum Computation of Molecular Linear Response Properties on a
  Superconducting Quantum Processor. \emph{The Journal of Physical Chemistry
  Letters} \textbf{2022}, \emph{13}, 9114--9121, PMID: 36154018\relax
\mciteBstWouldAddEndPuncttrue
\mciteSetBstMidEndSepPunct{\mcitedefaultmidpunct}
{\mcitedefaultendpunct}{\mcitedefaultseppunct}\relax
\EndOfBibitem
\bibitem[Temme \latin{et~al.}(2017)Temme, Bravyi, and Gambetta]{temme2017error}
Temme,~K.; Bravyi,~S.; Gambetta,~J.~M. Error mitigation for short-depth quantum
  circuits. \emph{Physical review letters} \textbf{2017}, \emph{119},
  180509\relax
\mciteBstWouldAddEndPuncttrue
\mciteSetBstMidEndSepPunct{\mcitedefaultmidpunct}
{\mcitedefaultendpunct}{\mcitedefaultseppunct}\relax
\EndOfBibitem
\bibitem[Endo \latin{et~al.}(2018)Endo, Benjamin, and Li]{endo2018practical}
Endo,~S.; Benjamin,~S.~C.; Li,~Y. Practical quantum error mitigation for
  near-future applications. \emph{Physical Review X} \textbf{2018}, \emph{8},
  031027\relax
\mciteBstWouldAddEndPuncttrue
\mciteSetBstMidEndSepPunct{\mcitedefaultmidpunct}
{\mcitedefaultendpunct}{\mcitedefaultseppunct}\relax
\EndOfBibitem
\end{mcitethebibliography}


\providecommand{\latin}[1]{#1}
\makeatletter
\providecommand{\doi}
  {\begingroup\let\do\@makeother\dospecials
  \catcode`\{=1 \catcode`\}=2 \doi@aux}
\providecommand{\doi@aux}[1]{\endgroup\texttt{#1}}
\makeatother
\providecommand*\mcitethebibliography{\thebibliography}
\csname @ifundefined\endcsname{endmcitethebibliography}
  {\let\endmcitethebibliography\endthebibliography}{}
\begin{mcitethebibliography}{4}
\providecommand*\natexlab[1]{#1}
\providecommand*\mciteSetBstSublistMode[1]{}
\providecommand*\mciteSetBstMaxWidthForm[2]{}
\providecommand*\mciteBstWouldAddEndPuncttrue
  {\def\EndOfBibitem{\unskip.}}
\providecommand*\mciteBstWouldAddEndPunctfalse
  {\let\EndOfBibitem\relax}
\providecommand*\mciteSetBstMidEndSepPunct[3]{}
\providecommand*\mciteSetBstSublistLabelBeginEnd[3]{}
\providecommand*\EndOfBibitem{}
\mciteSetBstSublistMode{f}
\mciteSetBstMaxWidthForm{subitem}{(\alph{mcitesubitemcount})}
\mciteSetBstSublistLabelBeginEnd
  {\mcitemaxwidthsubitemform\space}
  {\relax}
  {\relax}

\bibitem[Peruzzo \latin{et~al.}(2014)Peruzzo, McClean, Shadbolt, Yung, Zhou,
  Love, Aspuru-Guzik, and O’Brien]{peruzzo2014variational}
Peruzzo,~A.; McClean,~J.; Shadbolt,~P.; Yung,~M.-H.; Zhou,~X.-Q.; Love,~P.~J.;
  Aspuru-Guzik,~A.; O’Brien,~J.~L. A variational eigenvalue solver on a
  photonic quantum processor. \emph{Nat. Commun.} \textbf{2014}, \emph{5}\relax
\mciteBstWouldAddEndPuncttrue
\mciteSetBstMidEndSepPunct{\mcitedefaultmidpunct}
{\mcitedefaultendpunct}{\mcitedefaultseppunct}\relax
\EndOfBibitem
\bibitem[Anand \latin{et~al.}(2022)Anand, Schleich, Alperin-Lea, Jensen, Sim,
  D^^c3^^adaz-Tinoco, Kottmann, Degroote, Izmaylov, and
  Aspuru-Guzik]{anand2022}
Anand,~A.; Schleich,~P.; Alperin-Lea,~S.; Jensen,~P. W.~K.; Sim,~S.;
  D^^c3^^adaz-Tinoco,~M.; Kottmann,~J.~S.; Degroote,~M.; Izmaylov,~A.~F.;
  Aspuru-Guzik,~A. A quantum computing view on unitary coupled cluster theory.
  \emph{Chem. Soc. Rev.} \textbf{2022}, \emph{51}, 1659--1684\relax
\mciteBstWouldAddEndPuncttrue
\mciteSetBstMidEndSepPunct{\mcitedefaultmidpunct}
{\mcitedefaultendpunct}{\mcitedefaultseppunct}\relax
\EndOfBibitem
\bibitem[Jordan and Wigner(1928)Jordan, and Wigner]{Jordan1928}
Jordan,~P.; Wigner,~E. {\"U}ber das Paulische {\"A}quivalenzverbot.
  \emph{Zeitschrift f{\"u}r Physik} \textbf{1928}, \emph{47}, 631--651\relax
\mciteBstWouldAddEndPuncttrue
\mciteSetBstMidEndSepPunct{\mcitedefaultmidpunct}
{\mcitedefaultendpunct}{\mcitedefaultseppunct}\relax
\EndOfBibitem
\end{mcitethebibliography}

\end{document}

\section{Second-order partial derivative with respect to orbital parameters}
The second-order partial derivative of the OO-VQE cost function $E(\bmf, \bmkp, \bmth)$ with respect to $\bmkp$ is 
\begin{align}
\pdv{E(\bmf, \bmkp, \bmth)}{\kappa_{pq}}{\kappa_{rs}} \big|_{\bm{\kappa}=\bm{0}} 
& = \frac{1}{2} 
\ev{\left[ \left[ \hat{H}_\mr{full}(\bmf, \bmkp), \hat{\kappa}_{rs}, \right], \hat{\kappa}_{pq} \right]}{\Psi} \big|_{\bm{\kappa}=\bm{0}} \nonumber \\
& + \frac{1}{2} 
\ev{ \left[ \left[ \hat{H}_\mr{full}(\bmf, \bmkp), \hat{\kappa}_{pq} \right], \hat{\kappa}_{rs} \right]}{\Psi} \big|_{\bm{\kappa}=\bm{0}}. 
\end{align}
The following equation holds:
\begin{align*}
& \ev{\left[ \left[ \hat{H}_\mr{full}(\bmf, \bmkp), \hat{\kappa}_{pq}, \right], \hat{\kappa}_{rs} \right]}{\Psi} \big|_{\bm{\kappa}=\bm{0}} \\
= & \left( - h_{ps}\rho^{(1)}_{rq} - h_{qr}\rho^{(1)}_{ps} + \delta_{qr} \sum_{t} h_{tp} \rho^{(1)}_{ts} + \delta_{ps} \sum_{t} h_{qt} \rho^{(1)}_{rt}  \right.\\
 & + \delta_{ps} \sum_{tuv} g_{qtuv} \rho^{(2)}_{rtuv} + \delta_{qr} \sum_{tuv} g_{tupv} \rho^{(2)}_{tusv}
 + \sum_{tu} g_{qstu} \rho^{(2)}_{prtu} + \sum_{tu} g_{tupr} \rho^{(2)}_{tuqs} \\
 & \left. - \sum_{tu} g_{qtru} \rho^{(2)}_{ptsu} - \sum_{tu} g_{qtur} \rho^{(2)}_{ptus}
 - \sum_{tu} g_{stpu} \rho^{(2)}_{rtqu} - \sum_{tu} g_{tspu} \rho^{(2)}_{trqu} \right) \\
 & - \left( p \leftrightarrow q \right) - \left( r \leftrightarrow s \right) + \left( p \leftrightarrow q \text{ and } r \leftrightarrow s \right), 
\end{align*}
where $\left( p \leftrightarrow q \right)$ indicates that the terms in the first parenthesis in the right hand side of the equation are repeated with swapping indices $p$ and $q$.
Therefore, one can calculate $\pdv{E(\bmf, \bmkp, \bmth)}{\kappa_{pq}}{\kappa_{rs}}$ on classical computers by using the values of 1,2-RDMs.

\section{Molecular geometries used in numerical simulations}
The molecular geometries used in Sec. 4.2 and Sec. 5.2 in the main text are summarized in Table~\ref{tab: geometries}.

\begin{table*}[] 
 \caption{Geometries of molecules. ``$(\mr{X}, (x,y,z))$" denotes three dimensional coordinates $x,y,z$ of atom X in units of Bohr. 
 \label{tab: geometries}
 }
 \begin{tabular}{c|l}
 \hline \hline
 Molecule & Geometry  \\ \hline
 \ce{H2O} & (O, (0, 0, 0.20189834)), (H, (0, 1.48369957, -0.80757446)), \\
 & (H, (0, -1.48369957, -0.80757446)) \\
 \ce{C4H4S} &
(S, (0, 0, 2.32317)), \\
& (C, (0,  2.37947, -0.080994)),
(C, (0, -2.37947, -0.080994)), \\
& (C, (0,  1.36487, -2.40628)),
(C, (0, -1.36487, -2.40628)), \\
& (H, (0,  4.32679,  0.43743)),
(H, (0, -4.32679,  0.43743)), \\
& (H, (0,  2.47276, -4.09925)),
(H, (0, -2.47276, -4.09925)) \\
\ce{C4H4O} & 
(O, (0, 0, 2.17507477)), \\
& (C, (0,  2.09381655, 0.64439661)),
(C, (0, -2.09381655, 0.64439661)), \\
& (C, (0,  1.36627199, -1.78957064)),
(C, (0, -1.36627199, -1.78957064)), \\
& (H, (0,  3.87771801,  1.56847268)), 
(H, (0, -3.87771801,  1.56847268)), \\
& (H, (0,  2.57002753, -3.40528648)),
(H, (0, -2.57002753, -3.40528648))
\\ \hline \hline
 \end{tabular}
\end{table*}

\section{Unitary coupled-cluster singles and doubles ansatz}
In the numerical simulations in the main text, we employed unitary coupled-cluster singles and doubles ansatz~\cite{peruzzo2014variational, anand2022} with the first-order Trotterization.
We defined the ansatz in the fermionic form as
\begin{align}
 \ket{\psi_\mr{UCCSD}^\mr{(fermion)}(\bmth)} &  =\hat{U}_\mr{UCCSD}^\mr{(fermion)}(\bmth) \ket{\mr{HF}},\\
 \hat{U}_\mr{UCCSD}^\mr{(fermion)}(\bmth) &= e^{\hat{T}_s - \hat{T}_s^\dag} e^{\hat{T}_{d1} - \hat{T}_{d1}^\dag} e^{\hat{T}_{d2} - \hat{T}_{d2}^\dag}, \\
 \hat{T}_s &= \sum_{o,v} \theta_{ov}^{(s)} \left(  \hat{a}_{v\uparrow}^\dag \hat{a}_{o\uparrow} + \hat{a}_{v\downarrow}^\dag \hat{a}_{o\downarrow} \right), \\
 \hat{T}_{d1} &= \sum_{o, v} \theta^{(d1)}_{ov} \hat{a}_{v\uparrow}^\dag \hat{a}_{o\uparrow} \hat{a}_{v\downarrow}^\dag \hat{a}_{o\downarrow}, \\
 \hat{T}_{d2} &= \sum_{(o_1, v_1) \neq (o_2, v_2)} \sum_{\sigma, \tau = \uparrow, \downarrow} \theta_{o_1 o_2 v_1 v_2}^{(d2)} \hat{a}_{v_1\sigma}^\dagger \hat{a}_{o_1\sigma} \hat{a}_{v_2\tau}^\dagger \hat{a}_{o_2\tau},
\end{align}
where $\ket{\mr{HF}}$ is the Hartree-Fock state and $o,o_1,o_2 (v,v_1,v_2)$ are the occupied (virtual) orbitals in the Hartree-Fock state.
The circuit parameters $\bmth$ of this ansatz are $\{\theta_{ov}^{(s)} \} \cup \{\theta_{ov}^{(d1)} \} \cup \{\theta_{o_1o_2v_1v_2}^{(d2)} \}$.
To express this ansatz state in the qubit form, we performed the Jordan-Wigner transformation~\cite{Jordan1928} for $\hat{T}_s - \hat{T}_s^\dag, \hat{T}_{d1} - \hat{T}_{d1}^\dag, \hat{T}_{d2} - \hat{T}_{d2}^\dag$ and $\ket{\mr{HF}}$.
For example, if $\hat{T}_s - \hat{T}_s^\dag$ is transformed into the sum of multiqubit Pauli operators as $i\sum_{ov} \theta_{ov}^{(s)} \sum_k \alpha_k^{(ov)} \hat{P}_k^{(ov)}$, where $\alpha_k^{(ov)} \in \mathbb{R}$ and $\hat{P}_k^{(ov)}$ is a multiqubit Pauli operator, we (approximately) implement $e^{\hat{T}_s - \hat{T}_s^\dag}$ by the first-order Trotterization:
\begin{equation}
 \prod_{ov} \prod_k \exp[i \alpha_k^{(ov)} \theta_{ov}^{(s)} \hat{P}_k^{(ov)}].
\end{equation}
The operators $e^{\hat{T}_{d1} - \hat{T}_{d1}^\dag}$ and $e^{\hat{T}_{d2} - \hat{T}_{d2}^\dag}$ were approximately implemented as qubit operators in the same way.
We applied these approximated qubit representation of operators to the qubit representation of $\ket{\mr{HF}}$ and used it as the ansatz in the numerical calculation in the main text.

\section{Estimating expectation values of observables on quantum computers}
In this section, we explain how to evaluate expectation values of observables on quantum computers and the procedures of numerical simulations in Sec. 5.2 in the main text.

Let us consider evaluating an expectation value $\ev{\hat{O}}{\psi}$ for an observable $\hat{O}$ and a state $\ket{\psi}$ on $N$ qubits.
To evaluate $\ev{\hat{O}}{\psi}$ on quantum computers, we decompose $\hat{O}$ as
\begin{equation} \label{eq: O decomp}
 \hat{O} = \sum_{n=1}^{M} c_n \hat{P}_n,
\end{equation}
where $\hat{P}_n$ is the Pauli operator on $N$ qubits (i.e., $\hat{P}_n \in \{I,X,Y,Z\}^{\otimes N}$ with $X,Y,Z$ being the Pauli matrix on a single qubit) and $M$ is the number of Pauli operators contained in $\hat{O}$.
In the most naive way to evaluate $\ev{\hat{O}}{\psi}$, we perform the projective measurement of $\hat{P}_n$ to the state $\ket{\psi}$ for all $n=1,\cdots,M$.
For each projective measurement of $\hat{P}_n$, we repeatedly prepare $\ket{\psi}$ and run specific quantum circuit, which yields the eigenvalues $\pm 1$ of $\hat{P}_n$ (note that $\hat{P}_n^2 = \hat{I}$ implies the eigenvalues are $\pm 1$).
By averaging the outputs of $\pm 1$, we estimate the expectation value $\ev{\hat{P}_n}{\psi}$.
The whole expectation value $\ev{\hat{O}}{\psi}$ is estimated by calculating
\begin{equation}
 \ev{\hat{O}}{\psi} = \sum_{n=1}^M c_n \ev{\hat{P}_n}{\psi}
\end{equation}
with classical computers.

In this procedure, the number of executions of the quantum circuits to obtain the eigenvalues of $\hat{P}_n$ is called the number of measurements or the number of {\it shots}.
The difference between the estimated expectation values and the exact expectation value ($\ev{\hat{O}}{\psi}$) gets smaller when we increase the number of shots.
Although the quantum circuit execution described above is completely parallelizable when one has a lot of quantum computers, it is strongly desirable to reduce the number of shots for evaluating expectation values to save the time for computing.

In the numerical simulations in Sec. 5.2 in the main text, we simulated the above procedure with classical computers by randomly generating the results of the projective measurements to evaluate the expectation values of the observables that appear in the formulas of the analytical and numerical derivatives.

\bibliography{bibliography}